\documentclass[aps,nofootinbib,prx,twocolumn,notitlepage,floatfix,superscriptaddress,10pt,tightenlines]{revtex4-2}
\usepackage[english]{babel} 
\usepackage{xcolor}
\usepackage{amsfonts,amsmath,amssymb,mathtools,bm}
\usepackage{setspace}
\usepackage{xspace}
\usepackage{newfloat,algcompatible}
\usepackage{etoolbox}
\usepackage{bbold}
\usepackage{enumitem}
\AtBeginEnvironment{algorithm}{\noindent\hrulefill\par\nobreak\vskip-5pt}
\usepackage{newfloat}
\usepackage{wrapfig}

\DeclareFloatingEnvironment[
    fileext=loa,
    listname=List of Algorithms,
    name=ALGORITHM,
    placement=tbhp,
]{algorithm}

\usepackage{acronym,array,bm,color,dcolumn}
\usepackage{epsfig,empheq}
\usepackage{nomencl,revsymb4-1,upgreek,url}
\usepackage{graphicx}
\graphicspath{{figures/}}
\usepackage[pdfusetitle]{hyperref}
\usepackage{xspace}
\usepackage{dcolumn}
\usepackage{listings}

\usepackage{soul}
\PassOptionsToPackage{caption=false}{subfig}
\hypersetup{colorlinks=true}
\hypersetup{citecolor={blue}, colorlinks={true}, filecolor={blue},
    linkcolor={blue}, pdfkeywords= pdfsubject=, urlcolor={blue}}

\newcommand*{\ket}[1]{\ensuremath{\left\vert{#1}\right\rangle}}

\newcommand{\prob}{\mathsf{Pr}}
\newcommand{\CZ}{\ensuremath{\mathsf{CZ}}\xspace}
\newcommand*{\gx}{\ensuremath{\mathsf{G}_x}}
\newcommand*{\gxone}{\ensuremath{\mathsf{G}_{x,1}}}
\newcommand*{\gxtwo}{\ensuremath{\mathsf{G}_{x,2}}}
\newcommand*{\gy}{\ensuremath{\mathsf{G}_y}}

\newcommand*{\ux}{\ensuremath{U_{\mathsf{G}_x}}}
\newcommand*{\uy}{\ensuremath{U_{\mathsf{G}_y}}}

\newcommand*{\abs}[1]{\left\vert #1 \right\vert}
\newcommand*{\expect}[1]{\ensuremath{\left\langle{#1}\right\rangle}}
\newcommand*{\expectation}[1]{\ensuremath{\mathbb{E}\left[#1\right]}}

\newcommand*{\etaopt}[1][]{\eta_{{{\rm opt}\if\relax\detokenize{#1}\relax\else,#1\fi}}}
\DeclareMathOperator*{\E}{\mathbb{E}}
\newcommand*{\gain}{\ensuremath{g}}

\begin{document}

\title{Fast-feedback protocols for calibration and drift control in quantum computers}
\author{Alicia B. Magann}\thanks{These two authors contributed equally.}
\affiliation{Quantum Performance Laboratory, Sandia National Laboratories, Albuquerque, NM 87123}

\author{Nathan E. Miller}\thanks{These two authors contributed equally.}
\affiliation{Quantum Performance Laboratory, Sandia National Laboratories, Albuquerque, NM 87123}
\affiliation{Department of Electrical and Computer Engineering, Georgia Institute of Technology, Atlanta, GA 30332}
\affiliation{MIT Lincoln Laboratory, Lexington, MA 02421}

\author{Robin Blume-Kohout}
\affiliation{Quantum Performance Laboratory, Sandia National Laboratories, Albuquerque, NM 87123}

\author{Peter Maunz}
\affiliation{Quantum Performance Laboratory, Sandia National Laboratories, Albuquerque, NM 87123}
\affiliation{Currently at IonQ, College Park, Maryland 20740, USA}

\author{Kevin C. Young}
\email[Email address: ]{kyoung@sandia.gov}
\affiliation{Quantum Performance Laboratory, Sandia National Laboratories, Livermore, CA 94708}
\date{\today}

\begin{abstract}

\noindent We introduce two classes of lightweight, adaptive calibration protocols for quantum computers that leverage fast feedback. The first enables shot-by-shot updates to device parameters using measurement outcomes from simple, indefinite-outcome quantum circuits. This low-latency approach supports rapid tuning of one or more parameters in real time to mitigate drift. The second protocol updates parameters after collecting measurements from definite-outcome circuits (e.g.~syndrome extraction circuits for quantum error correction), balancing efficiency with classical control overheads.  We use numerical simulations to demonstrate that both methods can calibrate 1- and 2-qubit gates rapidly and accurately even in the presence of decoherence, state preparation and measurement (SPAM) errors, and parameter drift. We propose and demonstrate effective adaptive strategies for tuning the hyperparameters of both protocols.  Finally, we demonstrate the feasibility of real-time \emph{in-situ} calibration of qubits performing quantum error correction, using only syndrome data, via numerical simulations of syndrome extraction in the $[[5,1,3]]$ code.

\end{abstract}

\maketitle

\section{Introduction}

Future quantum computers are expected to contain millions of qubits that each require precisely calibrated state preparation, gate, and measurement operations. These operations will be influenced by a long list of experimentally tunable parameters. Depending on the nature of the qubits, control parameters may include applied magnetic field strength, the amplitudes, phases, frequencies, and polarizations of laser or microwave pulses, or the shape of voltage pulses applied to surface electrodes. When these parameters deviate from their ideal values, the corresponding operations will cause more errors and the quantum computer's performance will degrade.  So parameters need to be \textit{calibrated}. Calibrating quantum computers by hand is already impractical and will become impossible on future utility-scale machines.  Therefore, realizing the full potential of large-scale quantum computers will require deploying automated calibration tools that rapidly adjust parameters to performance-optimizing values, and then \textit{re}adjust parameters if they drift during operation, without increasing error rates or downtime too much.  We introduce and demonstrate such tools in this paper.

Today, most quantum computers are calibrated with high-latency \textit{batch calibration} protocols.  These include parameter scans \cite{Tornow_2022} and iterative estimation and update protocols \cite{PhysRevLett.112.240504,carignandugas2023errorreconstructioncompiledcalibration,PhysRevApplied.13.044071,PhysRevLett.112.240503,PhysRevA.91.052306,abbasgholinejad2023extremum,jeanette2025blindcalibrationquantumcomputer}.  Batch calibration protocols made sense when updating control parameters took a long time.  For example, if qubits are controlled with an arbitrary waveform generator, adjusting a single parameter can take several minutes.  In this formerly common scenario, it is desirable to collect as much data as possible between update rounds.  Batched calibration approaches can be highly suboptimal \cite{van2013quantum, PhysRevA.89.062321, PhysRevA.92.022326}, however, as they require sustained downtime, long intervals between recalibrations, and risk making adjustments based on data that may be several minutes old and reflect an outdated noise environment. Increasingly, quantum computers are controlled by systems that use field-programmable gate arrays (FPGAs). Modern FPGA control systems impose much less control latency and permit faster calibration and drift compensation \cite{Dumoulin_Stuyck_2024,berritta2025efficientqubitcalibrationbinarysearch}. Calibration protocols that leverage these new capabilities in an automated manner can enhance performance, minimize system downtime, and maximize computational capability.  

\begin{figure*}[t] 
\centering
\includegraphics[width=2\columnwidth]{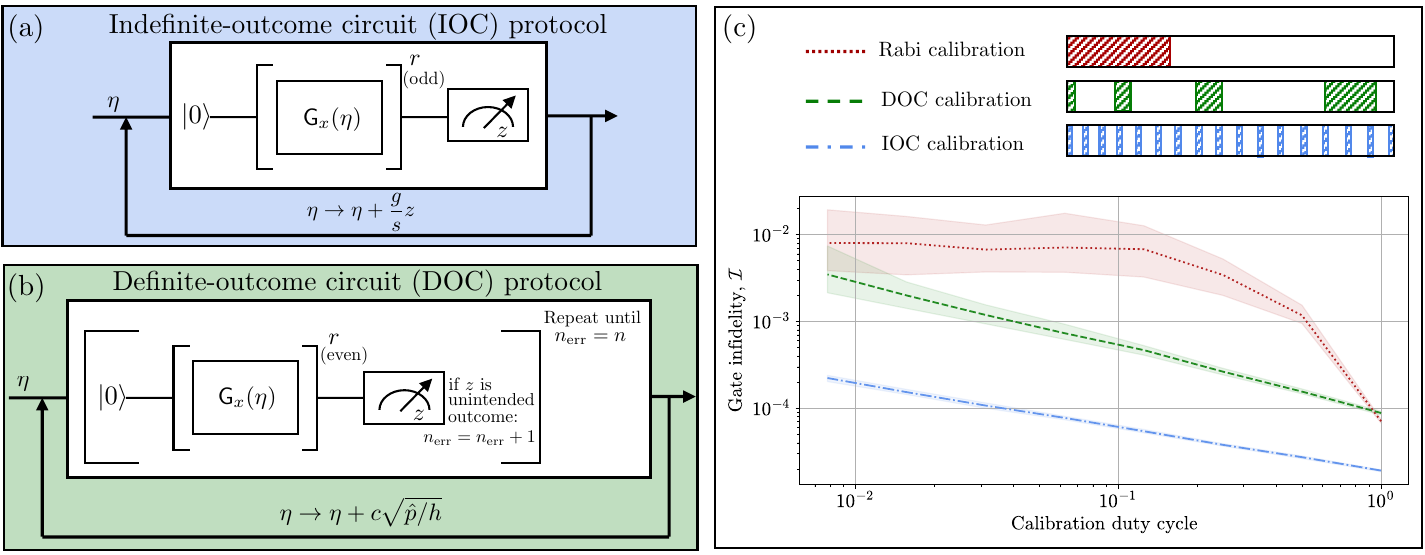} 
\caption{\textbf{IOC and DOC calibration protocols compared to Rabi calibration.} Many calibration protocols run batches of circuits, take many measurements, and require significant data processing to determine parameter updates. Such approaches are insensitive to short-time drift, and relying on them can reduce the amount of time available to run useful circuits on the quantum computer. Our proposed IOC and DOC protocols are diagrammed in panels (a) and (b), respectively, in the context of a simple example calibration of a $\gx$ gate. As shown in panel (c), these protocols can significantly outperform a batched Rabi curve-fitting calibration protocol in terms of tuned gate infidelity and experiment uptime. Further details of this simulation comparison can be found in App. \ref{App:TraditionalCal}.} 
\label{Fig:infidvsdutycycle}
\end{figure*}

We describe two families of fast-feedback calibration methods designed to take full advantage of low-latency control hardware. They update control parameters in real time, based on the outcomes of $O(1)$ shots of compatible quantum circuits. We refer to the first family of protocols as \emph{IOC protocols} because they utilize indefinite-outcome circuits. These protocols enable shot-by-shot control parameter updates to support rapid parameter tuning and low-latency drift compensation. The second family of protocols, \emph{DOC protocols}, operate with definite-outcome circuits. Given that syndrome extraction circuits have definite outcomes, DOC calibration protocols are compatible with execution \emph{in situ} during quantum error correction. Both IOC and DOC calibration protocols are designed to work under the relatively weak assumptions that (1) qubits are manipulated via controllable quantum operations, (2) the control parameters that determine gate behavior are known and adjustable, and (3) the system has already undergone initial calibration so that qubits are well-defined and the fidelities of state preparation and gate operations are already significantly greater than 50\%. In simulation studies, IOC and DOC calibration protocols outperform commonly-used batch protocols using the same duty cycle by a substantial margin. These results are presented in Fig.~\ref{Fig:infidvsdutycycle}, with further details provided in App.~\ref{App:TraditionalCal}.

\section{Tuning and Drift Control with Indefinite-Outcome Circuits}\label{Sec:IOC}

Our first class of protocols are \textit{single-shot}, and provide updated control parameter values after every individual measurement outcome of a circuit.  Because they use  {indefinite-outcome circuits} (IOCs) that yield random outcomes in the absence of errors, we call them {IOC protocols}. We define an indefinite-outcome quantum circuit as one whose ideal (error-free) outcome distribution is uniform over two or more possible measurement outcomes.  IOC circuits on a single qubit yield ``0'' or ``1'' with equal (50\%) probability.

We begin our discussion of IOC protocols with a motivating single-qubit, single-parameter calibration example in Sec.~\ref{sec:IOCsimpleexample}. Section \ref{Sec:LowLatencyDriftMitigation} shows how IOC protocols can compensate for time-dependent parameter drift. Section~\ref{Sec:IOCAdaptiveStrategies} presents adaptive heuristics for adjusting hyperparameters to achieve optimal performance and robustness to state preparation and measurement (SPAM) error. In Sec.~\ref{Sec:multiparameter} we show how to extend IOC protocols to multi-parameter and multi-qubit contexts. Additional details can be found in the Appendices: App.~\ref{App:IOC} analyzes and simulates IOC calibration and compares it to stochastic gradient descent \cite{SGD} and batch protocols, App~\ref{App:Sub:IOCHeuristics} explores hyperparameter heuristics, and App. \ref{App:Noise} examines performance under varied noise processes.

\subsection{Calibrating a rotation angle}\label{sec:IOCsimpleexample}

To motivate IOC calibration, consider the problem of calibrating a single-qubit rotation gate $\gx$ that should perform a unitary ${\pi/2}$ rotation about the Pauli $x$ ($\sigma_x$) axis,
\begin{equation}
    \ux=\exp\left(\frac{i}{2} \frac{\pi}{2} \sigma_x\right).
\end{equation}
We assume the gate is performed by a control pulse with a strength or amplitude parameter $\eta$. If $\eta$ is too strong or too weak at time $t$, the gate will over- or under-rotate the qubit's state by some amount $\delta_t$,
\begin{equation}
    \ux(\delta_t)=\exp\!\left(\frac{i}{2}\left(\frac{\pi}{2} + \delta_t\right) \sigma_x\right),
\end{equation}
where $\delta_t$ is proportional to the difference $\Delta\eta_t$ between the control parameter's \textit{actual} value $\eta_t$ and its \textit{ideal} value $\etaopt$,
\begin{equation}
    \delta_t = \alpha (\eta_t - \etaopt) = \alpha\Delta\eta_t.
\end{equation}
We assume the proportionality constant $\alpha$, which captures the relationship between the experimental control parameter (e.g., phase or strength of an applied control field) and the resulting gate rotation angle, is (approximately) known.

To calibrate the $\gx$ gate, we run a simple circuit $C$ that probes its behavior.  We use the notation
\begin{align}
\begin{split}
    C &= \left( \gx\right)^r \\
    &= (\;\underbrace{\gx, \gx, \gx, \ldots, \gx}_{r \text{ times}}\;),
\end{split}    
\end{align}
where $ \left( \gx\right)^r$ indicates $r$ consecutive applications of the $\gx$ gate, we require $r\equiv 1\mod{4}$, and every circuit begins with preparation in the $|0\rangle$ state and ends with measurement of the computational ($\sigma_z$) basis. In this notation, the gates in a circuit are applied sequentially from right to left, meaning that the gate on the far right is executed first and the leftmost gate is applied last. For now, we assume state preparation and measurement (SPAM) are perfect (although we will relax this assumption later and show that our protocols are robust to decoherence and SPAM error).

Because the over/under-rotation error we're calibrating commutes with $\gx$, this circuit \textit{amplifies} it, just like standard Rabi oscillation circuits. The circuit terminates with a measurement whose outcome is a time-dependent random variable $Z_t$ (its distribution depends on the time-dependent error parameter $\eta_t$) that takes values $\pm1$.  We use $z_t = \pm 1$ to denote a specific realization of $Z_t$ in a given experiment. 

Now, if $\Delta\eta_t=0$ then there is no rotation error, the circuit $C$ prepares the $|+i\rangle$ eigenstate of $\sigma_y$ and then measures $\sigma_z$, and so $Z_t$ is uniformly distributed over $\{+1,-1\}$.  This is why we call $C$ an \textit{indefinite-outcome} circuit.  But if $\Delta\eta_t \ne 0$, then $Z_t$'s distribution is biased:
\begin{equation}
\begin{aligned}
\prob(Z_t=\pm1) &= \frac{1}{2} \left(1\mp\sin(r\alpha\Delta\eta_t)\right)\\
\end{aligned}
\label{eq:probabilities}
\end{equation}
Assuming that $r\alpha\Delta\eta_t$ is small, we can approximate the r.h.s.~by its leading order Taylor expansion to get
\begin{equation}
\begin{aligned}
    \prob(Z_t=z_t)  
        &\approx \frac{1}{2} + s_{z_t} \Delta\eta_t,
\end{aligned}
\label{eq:probabilitieslinear}
\end{equation}
in terms of the \textit{outcome sensitivity} $s_{z_t}$, defined as 
\begin{align}
    s_{z_t}   &= \left.\frac{d}{d\eta_t} \prob(Z_t=z_t)\right\vert_{\Delta\eta_t=0}\\
        &= -\frac{1}{2} z_t\,\alpha\, r.
\end{align}
For the single-qubit, single-parameter case studied here, it will be convenient to define the \textit{circuit sensitivity}, 
\begin{equation}
    s=\abs{s_{z_t}}=\alpha r/2.
\end{equation}
The expectation value of the measurement, $\expect{z_t} \approx - 2s \Delta\eta_t$, directly reveals the error in the control parameter $\Delta\eta_t$ that we wish to calibrate away.  To perform traditional batch calibration, we would repeat this measurement many ($M$) times, estimate the error parameter's value with the empirical mean
\begin{equation}
\widehat{\Delta\eta} = \frac{1}{2 s M} \sum_{i = 1}^M z_i,
\end{equation}
and then modify the control parameter $\eta_t\rightarrow \eta_t - \hat\delta_t/\alpha$ to approximately (up to shot noise) eliminate the rotation error.  But collecting $M$ shots takes time, and by the time the correction is dialed in, $\Delta\eta_t$ may have changed.

We therefore take a different approach. We adjust $\eta$ after each individual sample, according to the rule 
\begin{equation}
\begin{aligned}
    \eta_{t+1}
    &= \eta_t + \frac{\gain}{s} \,z_t \\
     &= \eta_t - \frac{\gain}{s^2}s_{z_t},
\end{aligned}
\label{eq:IOCupdaterule}
\end{equation}
where $\gain \in [0,1/2)$ is a \textit{gain} parameter that determines the magnitude of the update. This procedure is depicted in Fig. \ref{Fig:infidvsdutycycle}(a).

In this protocol, $\Delta\eta_t$ evolves as a stochastic process because each update step depends on the random result of a quantum measurement. We can compute the mean, $\mu_t= \expectation{\Delta\eta_t}$, and variance, $\sigma_t^2 = \expectation{\Delta\eta_t^2}-\mu_t^2$, of this process. To do so, we assume (again) that the linear expansion of the probabilities in Eq.~\eqref{eq:probabilitieslinear} holds.  The mean obeys the difference equation
\begin{align}
    \mu_{t+1} &= (1-2\gain) \mu_t,
    \label{eq:mudifferenceequation}
\end{align}
whose solution is
\begin{equation}
    \label{eq:mean}
    \mu_{t} = (1-2\gain)^t \mu_0.
\end{equation}
This confirms that the update rule above produces the desired result -- the expected rotation error converges exponentially to zero, at a rate that scales with the gain $\gain$ but does not depend on the sensitivity $s$ of the probe circuit. 

\begin{figure}[t] 
\centering
\includegraphics[width=1\columnwidth]{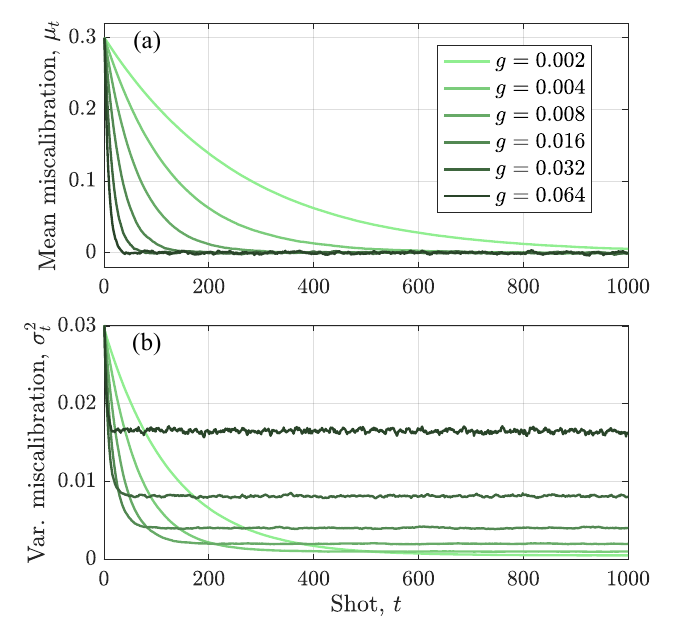} 
\caption{\textbf{IOC gain effects on convergence behavior.} Numerical simulations analyzing convergence behavior of the mean and variance of the miscalibration, $\Delta\eta_t$, as a function of the value of the IOC gain parameter, $g$, for calibrating the rotation angle of a $\gx$ gate as discussed in Sec. \ref{sec:IOCsimpleexample}. Results are computed over 10,000 realizations of the IOC protocol with $r=1$ and no drift. Different colors correspond to different values of $g$. Panel (a) presents the convergence of the mean, $\mu_t$, of the parameter deviation, $\Delta\eta_t$, as a function of the step, $t$, of the IOC protocol, when the initial offset is set to $\Delta\eta_0 = 0.3$ for all trajectories. We see that the rate of convergence increases with $g$ and all curves converge to $\mu_t\rightarrow 0$ for large $t$, as predicted by Eq. (\ref{eq:mean}). Panel (b) shows the convergence of the variance, $\sigma^2_t$ of the parameter deviation as a function of $t$ for initial offset sampled uniformly randomly from $\Delta\eta_0\in [-0.3, 0.3]$. Here, we observe that while the speed of convergence increases with $g$, the stationary value of $\sigma_t^2$ decreases with $g$. This tradeoff between rapid convergence and low stationary variance motivates the heuristics in Sec. \ref{Sec:IOCAdaptiveStrategies} for dynamically scheduling $g$ to be large at the outset, promoting rapid transient convergence, and then reduce over time to promote lower variance.}
\label{Fig:analyticalcomparisonmaintext}
\end{figure}

The variance of $\Delta\eta_t$ does not converge to zero, because for any finite gain $\gain$, shot noise (random measurement outcomes) causes $\Delta\eta_t$ to fluctuate around zero.  We compute the variance in Appendix~\ref{App:Sub:IOCBehavior}, and find that the asymptotic stationary variance of the rotation error is proportional to the gain parameter: 
\begin{equation}
    \sigma^2_\infty = \frac{\gain}{4s^2}
    \label{eq:stat_var_no_drift}
\end{equation}
We see that higher gain $\gain$ causes faster calibration (convergence of $\Delta\eta_t$ to zero), but also produces larger fluctuations once equilibrium is achieved. Figure \ref{Fig:analyticalcomparisonmaintext}(a) illustrates how increasing $\gain$ accelerates the convergence of the mean error to zero.  On the other hand, as shown in Figure \ref{Fig:analyticalcomparisonmaintext}(b), increasing $\gain$ makes the \textit{variance} of the error converge (faster) to a larger asymptotic value.

Large variance in $\eta_t$ is undesirable, because when the mean of $\eta_t$ is zero, the effective rate at which errors occur is proportional to its variance.  The variance can be reduced by using longer circuits, which have higher sensitivity $s$ (Eq.~\eqref{eq:stat_var_no_drift}).  However, excessively long circuits increase $s\Delta \eta_t$ so that, eventually, the linear expansion in Eq.~\eqref{eq:probabilitieslinear} becomes invalid.  This can cause the protocol fail by driving $\Delta\eta_t$ toward the wrong stable value -- i.e., one where $\sin(r\alpha\Delta\eta_t)=0$ not because $\Delta\eta = 0$, but because it's an integer multiple of $\frac{2\pi}{2\alpha}$.  We present heuristics for adaptively scheduling the gain and circuit depth parameters to accelerate convergence and minimize the asymptotic variance, while ensuring that the linear expansion in Eq.~\eqref{eq:probabilitieslinear} remains valid, in Sec.~\ref{Sec:IOCAdaptiveStrategies}.

\subsection{Low-latency drift mitigation}\label{Sec:LowLatencyDriftMitigation}

Any quantum computer calibration protocol can be used to mitigate drift in control parameters by repeating the calibration at regular intervals.  Full recalibration can be time-consuming, which motivates delaying recalibration to maximize device uptime. Doing so has a cost; long delays between recalibration events allow control parameters to drift, degrading both performance and reliability. In this section, we show that our IOC protocol can be used for effective, lightweight mitigation of drift.

Previously in Eqs. \eqref{eq:mudifferenceequation}-\eqref{eq:stat_var_no_drift}, we treated the ideal value of of the control parameter $\etaopt$ as a constant.  Now, to model drift, we promote $\etaopt$ to a stochastic process $\etaopt[t]$.  We analyze a variety of stochastic processes modeling different forms of drift in Appendix~\ref{App:Noise}, but to demonstrate the principle here we consider the specific example of a discrete-time, quasi-static random walk. We assume that $\etaopt$ stays fixed \textit{during} each shot of a quantum circuit, but from one shot to the next it changes as
\begin{equation}
    \etaopt[t] = \etaopt[t-1]+q_{t-1}\ell,
    \label{Eq:RandomWalkUpdate}
\end{equation}
where $\ell$ is the magnitude of the random walk at each step $t$ and $q_t\in\{-1,1\}$ is a random variable that dictates the sign of the step. 

\begin{figure} 
\centering
\includegraphics[width=\columnwidth]{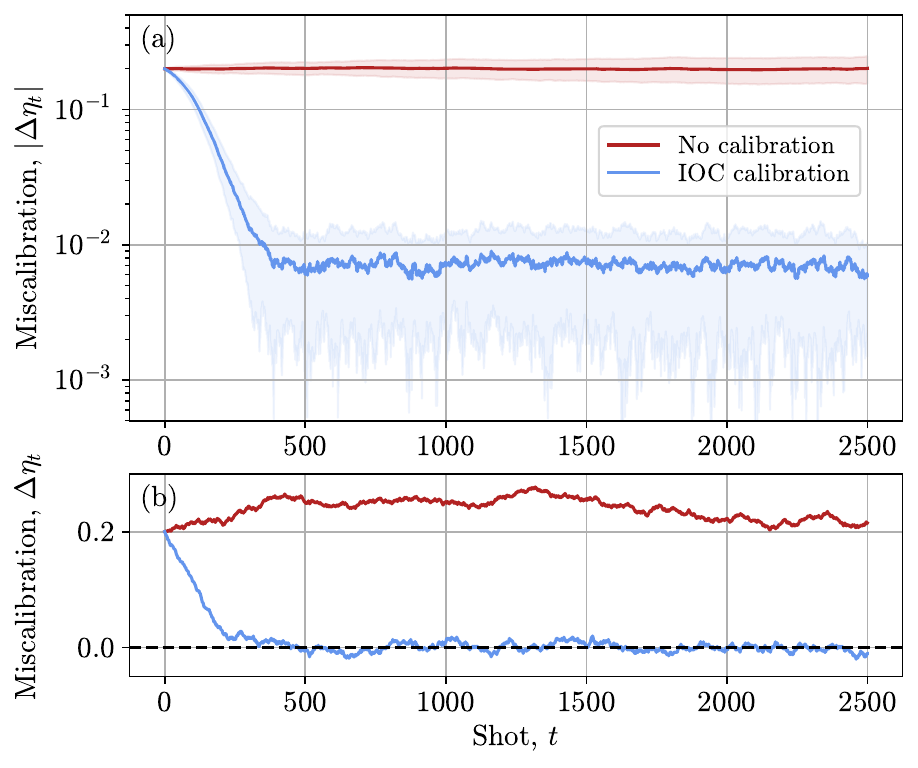} 
\caption{\textbf{Single-parameter IOC calibration for $\gx$ gate.} Results comparing the performance of the baseline {IOC} calibration protocol (blue) against an uncalibrated baseline (red). Here, the nominal value of the control parameter, $\etaopt$, drifts according to a discrete random walk, per Eq. (\ref{Eq:RandomWalkUpdate}), with $\ell=0.001$. The IOC circuit depth is set to $r=13$ to coherently amplify miscalibration errors for faster convergence, and the gain is set as $g=\ell s$. This simulation also incorporates per-gate depolarization with $p=0.001$ and depolarizing SPAM with $p_{SPAM}=0.01$. Panel (a) plots the magnitude of the miscalibration, $\Delta\eta_t$, as a function of shot, $t$, with $\Delta\eta_0=0.2$. The solid curve shows the mean and the shading shows associated standard deviation, computed over 50 trajectories. Panel (b), meanwhile, shows the miscalibration dynamics of a single trajectory. We observe that IOC calibration leads to rapid reduction and stabilization of the miscalibration error relative to the uncalibrated, drifting baseline.  }
\label{Fig:IOC}
\end{figure}

Our IOC calibration protocol can be used for drift mitigation without any modification.  We adjust the controllable parameter $\eta_t$ exactly as stated in the update rule in Eq.~\eqref{eq:IOCupdaterule}. Analysis shows that, as long as $\etaopt$ evolves according to an unbiased random walk, the mean $\mu_t=\expectation{\Delta\eta_t}$ evolves \textit{exactly} as it did for the undriven case in Eq.~\eqref{eq:mean} -- which means that the protocol successfully mitigates drift.  Figure \ref{Fig:IOC} illustrates the performance of the basic IOC protocol against this kind of random-walk noise.

Random-walk drift increases the value of the stationary variance somewhat (see Appendix \ref{App:Sub:IOCBehavior}), to:
\begin{equation}
    \sigma^2_\infty = \frac{g}{4s^2}+\frac{\ell^2}{4\gain}.
\end{equation}
Now the gain hyperparameter has \textit{two} competing effects -- it increases the ``calibration'' contribution to the stationary variance, but decreases the "drift'' term.  We can balance these effects and minimize the stationary variance by choosing $g=\ell s$, which makes the calibration step size $\gain/s$ equal in magnitude to the random walk step size $\ell$.  The optimal minimum stationary variance (and effective error rate) is therefore $\sigma_\infty^2 = \ell /2s$.

The discrete random walk model of drift is idealized, and in most cases we expect parameters to drift according to other laws.  So, in Appendix~\ref{App:Noise}, we analyze the performance of IOC calibration against a wide range of noise sources, including Ornstein-Uhlenbeck noise, $1/f$ noise, and jump processes, and we show that it works well for all of them. 

\subsection{Accelerated convergence and robustness to SPAM error}
\label{Sec:IOCAdaptiveStrategies}

As shown in Eq. (\ref{eq:mean}) and Fig. \ref{Fig:analyticalcomparisonmaintext}(a), choosing a large value for the gain parameter $\gain$ can accelerate transient behavior and lead to rapid convergence to a mean-zero process. But the variance of the stationary process grows with $\gain$, leading to short-lived but significant excursions away from the optimal parameters. In this section, we briefly motivate two potential heuristics for adaptive scheduling of the gain parameter and circuit sensitivity to promote rapid convergence, good stationary behavior, and robustness to drift. 

Gain and circuit scheduling can also play a crucial role in ensuring that the protocol remains robust against errors in state preparation and measurement (SPAM errors). For the protocol to operate reliably, any biases observed in the measurement results should arise primarily from calibration errors within the circuit itself, rather than from inaccuracies associated with SPAM (unless the protocol is being used specifically to calibrate SPAM operations). This can be enforced by choosing deep circuits so that gate errors dominate the overall error profile. In this section, we describe methods that implement this rule efficiently.

The presence of asymmetric SPAM errors can complicate the situation, but these errors can be effectively managed by implementing a strategy that alternates between different families of indefinite-outcome circuits. Ideally, these circuit families should be characterized by sensitivity parameters that are equal in magnitude but opposite in sign, specifically $s_z^{(1)}\approx -s_z^{(2)}$. This ensures that readout bias does not translate to bias in the inferred rotation error.  One practical approach to realize this alternating strategy involves distinguishing the two circuit families by the presence or absence of additional gates at the end of the circuit that swap the states $\ket{0}$ and $\ket{1}$. By alternating between these two configurations, the protocol can average out the biases that may be induced by asymmetric SPAM errors, leading to more reliable and accurate calibration outcomes.

In the simulations which follow, we incorporate SPAM error modeled as a depolarizing channel, as well as a per-gate depolarization error which occurs uniformly during the implementation of each gate in the circuit. A generic depolarization channel is described as:
\begin{equation}
    \rho \to p(I/d)+(1-p)\rho
\end{equation}
where $p$ is the depolarization probability and $I/d$ denotes the maximally mixed state, with $I$ denoting the identity operator on $n_q$ qubits and $d=2^{n_q}$ denoting the dimension of the state space. In our simulations, we take the depolarization probability associated with SPAM error, which we denote by $p_{SPAM}$, to be greater than the per-gate depolarization, denoted by $p$.

\subsubsection{Approximate error estimation}

The first gain and circuit scheduling heuristic is developed to promote rapid convergence to a low stationary variance of $\Delta\eta_t$. In particular, in App. \ref{App:Sub:IOCBehavior} we derive dynamical equations for the variance of $\Delta\eta_t$ under certain assumptions, and then seek a gain schedule that maximizes the rate of change of the variance. This results in an update rule for $\gain$ that depends on the mean and variance of $\Delta\eta_t$. The latter are then replaced by estimates that are computed from the measurement record, $\{z_t\}$, according to
\begin{equation}
    k_{t+1} = \frac{2\hat{\sigma}^2_t s^2}{1-4s^2\hat{\mu}^2_t},
\end{equation}
where $\hat{\mu}_t$ and $\hat{\sigma}^2_t$ denote the approximate estimates of the mean and variance of the error, $\Delta\eta_t$, respectively. These approximate error estimates can similarly be used to schedule the number of gate repetitions, $r$, in an effort to keep the estimated rotation error below some threshold value. In Appendix \ref{App:SubSub:ApproximateErrorEstimation}, we provide further details on this heuristic.

\subsubsection{Autocorrelation analysis}

The second heuristic offers an alternative prescription for hyperparameter tuning based on autocorrelation analysis. It is motivated by the fact that strong correlation in the measurement record $\{z_t\}$ indicates that the corrections to $\eta_t$ are not adequately compensating for the error, and thus the gain should be increased. On the other hand, anti-correlation in the measurement record implies that the feedback is causing oscillation (i.e., between over- and under-rotation in our simple, single-qubit example). In this case, the gain is too high and should be decreased. Accordingly, this heuristic for adaptively adjusting the gain, $\gain$, and number of circuit repetitions, $r$ proceeds by first initializing $\gain$ to be large, to promote fast initial convergence, and setting $r=1$. Then, the IOC protocol is initiated and after $h$ steps, the autocorrelation function of the last $h$ elements of the measurement record is estimated as $a = \sum_{t=1}^{h-1}z_tz_{t-1}$. If $a$ is large, this suggests that the measurement record is correlated and $\gain$ is increased; if it is small, this suggests anticorrelation and $\gain$ is decreased. For $a$ near zero, $r$ is increased. 
This protocol is particularly suited to lightweight implementations in FPGA or ASIC controllers due to its low computational overhead \cite{IOCImplementation}. We describe further details of this heuristic in \ref{App:SubSub:Autocorrelation}.

\subsection{Multi-Parameter Tuning and Drift Control}\label{Sec:multiparameter}

Here, we extend the IOC protocol to single- and multi-qubit settings with multiple tunable parameters. Specifically, we consider the task of tuning and stabilizing an $n$-qubit system with $m$ controllable parameters. After introducing the generalized theory, we present two examples calibrating and stabilizing: 1) phase and rotation-angle errors in a pair of single-qubit gates, and 2) the three nontrivial phases in a controlled-phase (\CZ) gate. 

In the single-qubit case described above, the IOC protocol utilizes circuits that ideally result in a uniform distribution over both possible outcomes $\ket0$ and $\ket1$. For the more general $n$-qubit IOC protocol, we define indefinite-outcome circuits as those for which the ideal outcome distribution is uniform over $M>1$ of the $2^n$ possible measurement outcomes, and zero otherwise. Examples of such circuits include those that create stabilizer states with at least one non-$Z$ stabilizer and terminate with a $Z$-basis measurement. 

Typically, multi-parameter IOC protocols require more than one circuit, $C$, which we collect into a set, $\{C^{(k)}\}$, indexed by $k$.  
We label outcomes as bitstrings, $z$, and denote the set of $m$ tunable parameters as $\bm{\eta}_t$. The probability of measuring outcome $z$ on circuit $C^{(k)}$ is denoted as $\prob(z\vert C^{(k)}, \bm{\eta})$. 

We assume knowledge of a control model that permits computation of derivatives of the outcome distribution with respect to $\bm{\eta}$, the tunable system parameters. We define the \textit{sensitivity vector} of a particular circuit-outcome pair as the gradient of the outcome's probability with respect to the control parameters, arranged as a column vector: 
\begin{align}
    \mathbf s^{(k)}_{z} = \mathbf\nabla_{\bm{\eta}} \prob(z\vert C^{(k)}, \bm{\eta}) 
        \big\vert_{\bm{\eta}=\bm{\eta}_{\rm{opt}}}
\end{align}
We can assemble these gradients into a Jacobian matrix:
\begin{equation}
\label{eq:multiparamjacobian}
    \mathcal{J} = \left(\bm{s}^{(1)}_0, \cdots \bm{s}^{(1)}_{2^n-1}, 
                        \bm{s}^{(2)}_0, \cdots \bm{s}^{(2)}_{2^n-1}, 
                        \cdots \right)^{\mathsf{T}}
\end{equation}
The Jacobian of the circuit set must be informationally complete. That is, we require sufficiently many circuits that $\mathsf{rank}(\mathcal{J}) = \mathsf{dim}(\bm{\eta})$. Additionally, we will see later that the algorithm will perform best if $\mathcal{J}$ has low condition number, so we can learn all of the parameters to approximately the same precision. The problem of identifying a set of circuits that satisfy these conditions bears strong similarities to the problem of experiment design in gate set tomography \cite{Nielsen_2021}. 

The multi-parameter IOC protocol proceeds at each step $t$ by first selecting and running a circuit  $C^{(k)}$. In the numerical examples below, we iterate through a predefined list of circuits, where each calibration round involves executing one shot of a single circuit from this list. More principled selection is certainly possible, but will likely depend on details of the calibration problem, and so we defer this to later research. After running the circuit, a computational basis measurement is performed that returns an $n$-bit string observation $z_t$. The tunable parameters are then adjusted using the generalized multiparameter update rule:
        \begin{equation}
            \bm{\eta}_{t+1} = \bm \eta_t - \frac{\gain}{\left\vert \bm{s}_{\bm{z}_t}^{(k)}\right\vert^2} \bm{s}_{\bm{z}_t}^{(k)}.
        \label{eq:multiparameterupdate}
        \end{equation}

\noindent \textbf{Example: One qubit, two parameters.} We consider two miscalibrated gates: a $\gx$ gate with some over- or under-rotation angle $\theta$, and a $\gy$ gate with the same over- or under-rotation angle $\theta$, and an additional phase shift $\phi$ (both with target rotation angles of $\pi/2$ around $\sigma_x$ and $\sigma_y$, respectively). This phase shift leads to a misalignment of the rotation axes, sometimes called a \textit{tilt error}.  The unitary operators associated with these gates are given by:
\begin{equation}
\begin{aligned}
    \ux(\theta) &= \exp\left(\frac{i}{2}\left(\frac{\pi}{2} + \theta\right) \sigma_x \right)
    \end{aligned}
\end{equation}
and
\begin{equation}
\begin{aligned}
    \uy(\theta, \phi) &= \exp\left( \frac{i}{2}\left(\frac{\pi}{2} + \theta\right) 
    \left(\sin(\phi) \,\sigma_x + \cos(\phi) \,\sigma_y\right) \right).
    \end{aligned}
\end{equation}
We assume that we have direct control over $\theta=\Delta\eta_{\theta}$ and $\phi=\Delta\eta_{\phi}$ via $\eta_{\theta}$ and $\eta_{\phi}$, respectively. We then define two IOC circuits that are sensitive to changes in the tunable parameters $\bm{\eta} = (\eta_\theta,\eta_\phi)$, given by:
\begin{align}
\begin{split}
    C^{(1)} &= \left(\gx,  \gy,  \gx,  \gy, \gx\right)\\ 
    C^{(2)} &= \left(\gx,  \gx,  \gy,  \gx,  \gy,  \gx,  \gy\right). 
\end{split}
\end{align}
We can directly calculate the sensitivity vectors for these circuits. 
\begin{align}
\begin{split}
    \mathcal{J} &= \left(\bm{s}^{(1)}_0, \bm{s}^{(1)}_1,
                        \bm{s}^{(2)}_0, \bm{s}^{(2)}_1 \right)^{\mathsf{T}} \\
                &= \begin{pmatrix}
                    0.316 & -0.316 & -0.588 & 0.588 \\
                    -0.632 & 0.632 & 0.392 & -0.392
                    \end{pmatrix}^{\mathsf{T}} 
\end{split}
\end{align}

\begin{figure} 
\centering
 \includegraphics[width=1\columnwidth]{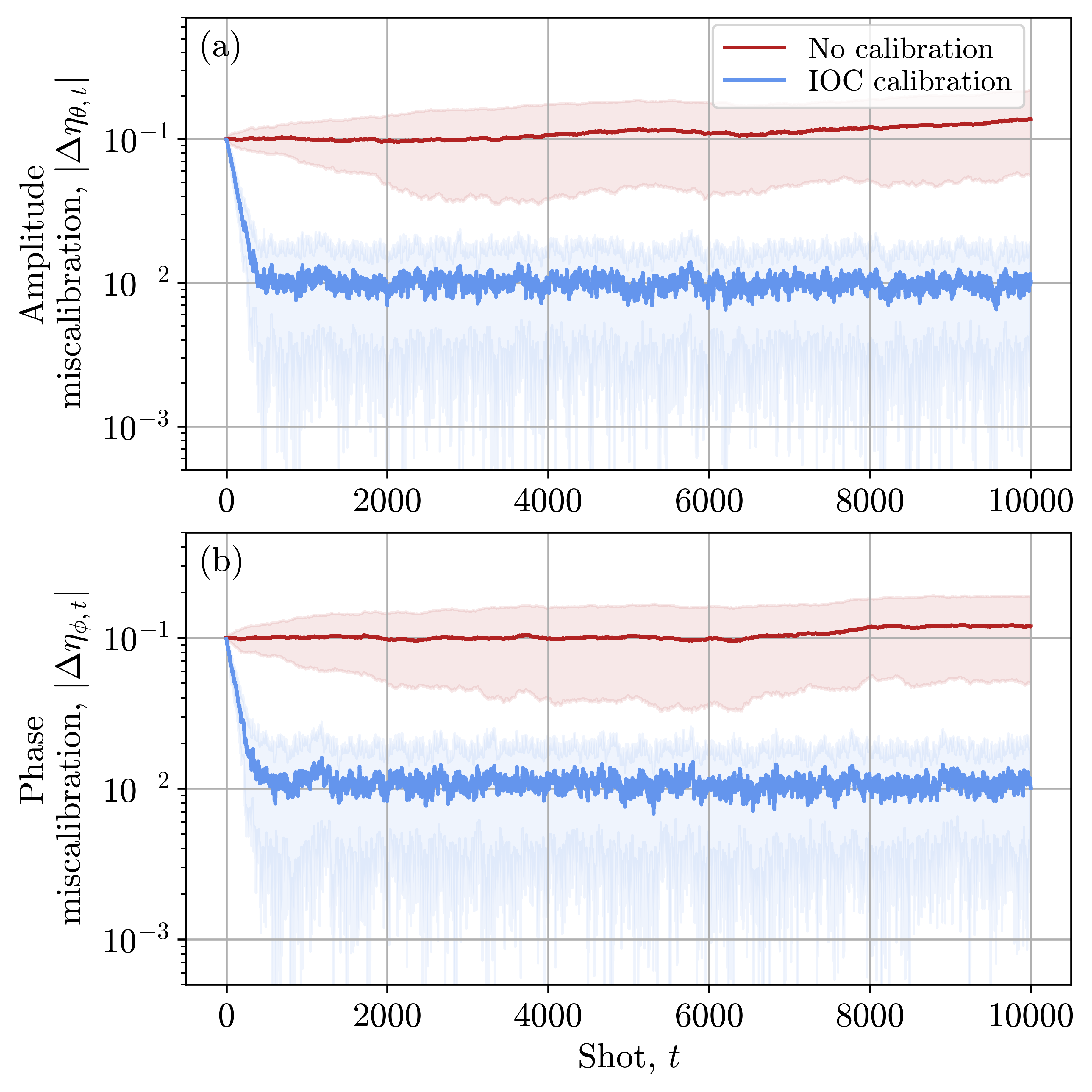} 
\caption{\textbf{Two-parameter IOC calibration for $\gx$ and $\gy$ gates.} Performance of the multiparameter IOC protocol applied to the two-parameter example described in Sec. \ref{Sec:multiparameter}. The magnitude of the parameter miscalibrations (a) $\Delta\eta_{\theta,t}$ and (b) $\Delta\eta_{\phi,t}$ are plotted as a function of shot, $t$ for simulations with IOC calibration (blue) and without any calibration (red). In these simulations, the nominal values, $\eta_{\theta,\text{opt}},\eta_{\phi,\text{opt}}$ of our tunable parameters each experience random walk drift with $\ell = 0.001$. These simulations also incorporate SPAM error with $p_{SPAM}=0.01$ and per-gate depolarization with $p=0.001$. The solid curves show the mean behavior over a set of 50 trajectories, and the shading shows the associated standard deviation. Both circuits are repeated $r=5$ times per calibration step to coherently amplify error rates with $g=0.001$. We observe that the multiparameter IOC protocol is successful in jointly tuning $\eta_\theta,\eta_\phi$ to reduce initial miscalibration error and maintain stable calibration in the presence of drift.}
\label{Fig:MultiParameterGxGy}
\end{figure}

Simulation results that include parameter drift, decoherence, and SPAM error are plotted in Fig. \ref{Fig:MultiParameterGxGy}. 

\vspace{.6cm}
\noindent \textbf{Example: Two qubits, three parameters.} In this example we consider a two-qubit controlled-phase (\CZ) gate suffering from phase errors:
\begin{align}
    U_{\CZ} =\exp
         &\left[  i \left(
              \left(\frac{\pi}{4}\right)I 
            + \left(\frac{\pi}{4} + \theta_{{ZZ}}\right) \sigma_z \otimes \sigma_z
        \right.\right.\\
     &  \notag
        \left.\left.
            -\left(\frac{\pi}{4} + \theta_{{IZ}}\right) I \otimes \sigma_z
            -\left(\frac{\pi}{4} + \theta_{{ZI}}\right) \sigma_z \otimes I  
        \right)\right]
\end{align}
This gate has three tunable parameters $\bm{\eta} = \left(\eta_{\theta_{{ZI}}}, \eta_{\theta_{{IZ}}}, \eta_{\theta_{{ZZ}}}\right)$, whose deviations from their nominal values are  $\theta_{{ZI}}$, $\theta_{{IZ}}$, and $\theta_{{ZZ}}$, respectively. When all the $\theta$ parameters are zero, the gate acts perfectly. 

We define two circuits that are jointly sensitive to changes in all three tunable parameters: 
\begin{align}
\begin{split}
    C^{(1)} &= \CZ, \gxtwo, \CZ, \gxtwo, \CZ, \gxtwo, H_{(1,2)}\\
    C^{(2)} &= \CZ, \gxone, \CZ, \gxone, \CZ, \gxone, H_{(1,2)}
\end{split}
\end{align}
For illustration purposes, we assume that the single-qubit operations, \gx$_{,i}$ (the $\gx$ gate acting on qubit $i$) and the Hadamard gate $H$, are perfect. When \CZ is perfectly calibrated, each of these circuits results in a uniform outcome distribution over all four two-bit strings. 

\begin{figure} 
\centering
 \includegraphics[width=1\columnwidth]{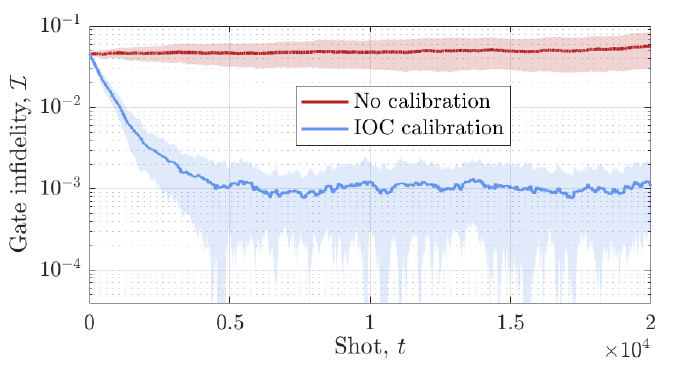} 
\caption{\textbf{Two-qubit, three-parameter IOC calibration for $\CZ$ gate.} Performance of the multiparameter IOC protocol, used here to stabilize three drifting phases of a two-qubit \CZ gate. Solid curves show mean \CZ gate infidelity as a function of shot, $t$, and shading shows associated standard deviation, taken over 50 realizations of the protocol. These simulations include drift in all three parameters, modeled as independent random walks that advance every shot with a step size of $\ell = 0.001$. Parameter offsets are initialized by setting $ \eta_{\theta_{{ZI}},opt,0}= \eta_{\theta_{{IZ}},opt,0}= \eta_{\theta_{{ZZ}},opt,0}=-0.25$. The gain used for the IOC protocol is $g = 2.5\times10^{-4}$ and $r=1$ repetition is used for each circuit. We see that IOC calibration is effective at jointly tuning the three control parameters to reduce and stabilize the $\CZ$ gate infidelity.}
\label{Fig:MultiParameter_CZ}
\end{figure}

The Jacobian for this experiment design is:
\begin{align}
\mathcal{J} &= \frac{1}{4}\left(
    \begin{array}{cccccccc}
        0 & 0 & 0 & 0 & 1 & 1 & -1 & -1 \\
        1 & -1 & 1 & -1 & 0 & 0 & 0 & 0 \\
        1 & -1 & -1 & 1 & 1 & -1 & -1 & 1
    \end{array}\right)^{\mathsf{T}}\\[-0.8em]
&\hphantom{\mathcal{J} = \frac{1}{4}\left(\right.}\underbrace{\hphantom{\begin{array}{cccc}1 & -1 & -1 & 1\end{array}}}_{C^{(1)}} 
\underbrace{\hphantom{\begin{array}{cccc}1 & 1 & -1 & -1\end{array}}}_{C^{(2)}}\notag
\end{align}
We have indicated the part of the Jacobian sensitive to each circuit, and we see that $C^{(1)}$ is sensitive to $\theta_{IZ}$ and $\theta_{ZZ}$, while $C^{(2)}$ is sensitive to $\theta_{{ZI}}$ and $\theta_{{ZZ}}$. The results of running the protocol are illustrated in Fig.~\ref{Fig:MultiParameter_CZ}. Rather than showing results for each of the three parameters, performance is captured here by a single metric: the gate infidelity. Throughout this paper, we use the term gate infidelity to refer to the entanglement infidelity \cite{PRXQuantum.6.030202}, which reduces here for the case of two unitaries to
\begin{equation} \label{eq:unitaryinfidelity}
    \mathcal{I} = 1-\frac{1}{d^2}\left\vert  
        \mathsf{Tr}\left[ 
             W^\dagger V 
            \right]\right\vert^2,
\end{equation} 
where the dimension $d=4$, $W = U_{\CZ}(\theta_{ZI}, \theta_{IZ}, \theta_{ZZ})$, and $V =  U_{\CZ}(0,0,0)$.

\section{Definite-Outcome Circuit Tuning and Drift Control}\label{sec:DOC}

We now turn to a second, distinct class of \textit{DOC protocols} that perform on-the-fly calibration using measurement results from \emph{definite-outcome circuits} (DOCs). Rather than adjusting control parameters after individual shots, DOC protocols can make adjustments every time an unexpected (``erroneous'') measurement result is observed from a definite-outcome circuit.  Definite-outcome circuits are ones that, when perfectly implemented, yield a unique deterministic outcome.  For a single qubit, they should produce either ``0'' or ``1'' deterministically. Definite-outcome circuits are common and very important in quantum computing -- examples include individual $\pi$ (Pauli) gates, $XY$-Ramsey sequences, and (most importantly) syndrome extraction circuits for quantum error correction. 

The DOC protocol repeatedly runs a DOC circuit until some number of failures are observed, and uses those observations to estimate the magnitude of the parameter error. The \textit{sign} of the error is undetermined, and the protocol works by alternately assuming that the sign of the estimated error is positive or negative. These alternating adjustments are self-correcting, so if a parameter update is made that worsens the error, it will then be corrected on the next round. 

Sec.~\ref{sec:DOCsimpleexample} outlines a simple, single-parameter DOC calibration example. In Sec.~\ref{sec:maintextadaptiveDOC} we motivate approaches for accelerating convergence and robustness of DOC calibration through adaptive scheduling heuristics. We then discuss the challenges facing multi-parameter extensions of DOC protocols in Sec. \ref{sec:multiparameterDOC}.  In Sec.~\ref{sec:DOCqec} we conclude by showing that DOC calibration methods can be applied to stabilizing systems running quantum error correction. In Apps. \ref{App:Sub:DOCHeuristic} and \ref{App:Noise}, we provide full details of our heuristic for accelerating convergence in DOC calibration and providing robustness to error sources, and show a variety of additional numerical simulation results.

\subsection{Simple Example}\label{sec:DOCsimpleexample}

We begin our discussion of DOC calibration with a simple example. In fact, we'll use the same gate and error model we considered when introducing the IOC calibration protocols in Sec. \ref{sec:IOCsimpleexample}, but applied to a different family of circuits. To calibrate the $\gx$ gate, we again use circuits comprising repeated $\gx$ gates:
\begin{align}
\begin{split}
    C &= \left( \gx\right)^r
\end{split}    
\end{align}
but this time we demand that $r$ is even,
thus enforcing that the circuit outcome is deterministic in the absence of noise. We continue with the convention that circuits begin with qubits prepared in the all-zeros state and end with a computational basis measurement. 

Again $Z_t:\{-1,+1\}$ is the random variable representing the outcomes of a circuit's measurement at time $t$, and $z_t$ is its specific realization. If $\Delta\eta_t=0$, the circuit will deterministically return $z_t=(-1)^{r/2}$. Because of this, circuits of this form are referred to as \textit{definite-outcome}. When the control parameter deviates from its ideal value, $\Delta\eta \ne 0$, the outcome distribution shifts:
\begin{equation}
\begin{aligned}
    \prob(Z_t=(-1)^{r/2} \vert C, \Delta\eta)  
        &= \cos^2\left(r \alpha \Delta\eta_t\right) \\
    \prob(Z_t=-(-1)^{r/2} \vert C, \Delta\eta)  
        &= \sin^2\left(r \alpha \Delta\eta_t\right) \\
\end{aligned}
\label{eq:probabilitiesDOC}
\end{equation}
Expanding to second order in $\Delta\eta$,
\begin{equation}
\begin{aligned}
    \prob(Z_t=(-1)^{r/2})  
        &\approx 1 - h \Delta\eta_t^2\\
    \prob(Z_t=-(-1)^{r/2})  
        &\approx h \Delta\eta_t^2
\end{aligned}
\label{eq:probabilitiesDOC2}
\end{equation}
Here we have defined the \textit{second-order outcome sensitivity}, 
\begin{align}
    h   &= \left.\frac{d^2}{d\eta^2} \prob(Z_t=(-1)^{r/2})\right\vert_{\Delta\eta=0}\\
        &= r^2 \alpha^2
\end{align}

The DOC calibration protocol aims to iteratively adjust $\eta$ in order to drive the probability of the unintended outcome $\prob(Z_t = -(-1)^{r/2})$ towards zero, ensuring that $\eta \approx \etaopt$. By repeatedly running the circuit $C$ until we observe some preselected number of failures, controlled by the cutoff parameter $n$, we can estimate $\prob(z_t = -(-1)^{r/2})$ as the empirical fraction $\hat p$ of unintended measurement outcomes. Combined with knowledge of the second-order sensitivity $h$, this enables an estimate\footnote{We adopt the maximum likelihood estimator here. In App.~\ref{App:DOC} we show evidence that it outperforms the minimum-variance unbiased estimator.} of $\abs{\Delta\eta_t}$ as 
\begin{equation}
    \label{eq:DOCdeltaeta}
    \widehat{\,\abs{\Delta\eta_t}\,} = \sqrt{\hat{p}/h}.
\end{equation}
At this point, we have estimated the size of the calibration step necessary to (approximately) eliminate the error, but we have learned nothing about the sign of the error! To proceed, we pick a calibration direction at random using a ``coin flip,'' $c\in\{1,-1\}$, and adjust the parameter as:
\begin{equation}
    \label{eq:DOCupdate}
    \eta \to \eta + c \sqrt{\hat{p}/h},
\end{equation} and then run the experiment again. This procedure is depicted in Fig. \ref{Fig:infidvsdutycycle}(b). 

If $\widehat{\,\abs{\Delta\eta_t}\,}$ worsens, then our initial correction was almost certainly in the wrong direction, prompting us to switch the sign of the coin variable $c\to-c$ for the next correction. Conversely, if the error improves, we can assume our previous calibration step was in the right direction. But we can't tell if we over-corrected or under-corrected, so we have no information about whether the next parameter update should be positive or negative. By consistently updating $c\to-c$, we maintain a uniform approach for all cases. This procedure is repeated in subsequent steps, alternating the sign of $c$ each time, until the estimate of the probability converges as desired. 

We can also choose to add a gain parameter to the update rule in Eq.~\eqref{eq:DOCupdate} as in the IOC protocol, but we have not seen a significant benefit to scaling this value in simulation. The sensitivity of the error estimation in the DOC protocol is controlled by the cutoff parameter, $n$. Choosing a larger value of the cutoff parameter reduces the effects of shot noise and improves the error estimates, but the expected number of shots required between calibration updates increases linearly with $n$. When choosing $r$, it is also important to ensure that $\hat{p}$ is likely to be a realistic value of $\hat{p}\leq1$ according to Eq.~\eqref{eq:DOCdeltaeta}. A large $r$ combined with a large $\Delta\eta$ will lead to a breakdown of the probability estimate, a regime in which the convergence behavior can become erratic.

\begin{figure}[t] 
\centering
\includegraphics[width=\columnwidth]{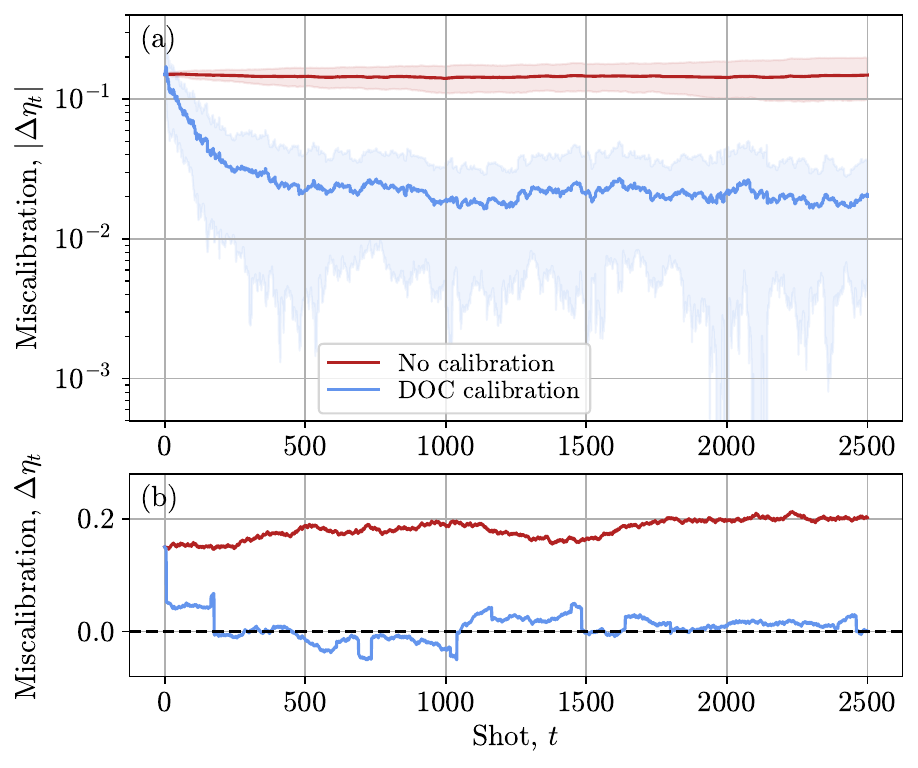} 
\caption{\textbf{Single-parameter DOC calibration for $\gx$ gate.} Results comparing the performance of the DOC protocol (blue) against an uncalibrated baseline (red). 
In these simulations, $\etaopt$ drifts according to a random walk with step size $\ell=0.001$. This simulation also incorporates per-gate depolarization with $p=0.001$ and depolarizing SPAM with $p_{SPAM}=0.01$. We set $r=6$ to coherently amplify miscalibration error, equivalent to a total target rotation of $3\pi$. Panel (a) shows the magnitude of the miscalibration, $\Delta\eta_t$, as a function of shot, $t$, starting from an initial condition of $\Delta\eta_0=0.15$. Solid curves and associated shading indicate the mean and standard deviation, respectively, computed over 50 trajectories. Panel (b) shows the dynamics of a single trajectory. The DOC protocol shown here is effective in quickly reducing the initial miscalibration offset, and then maintaining stable behavior in the presence of drift. As the miscalibration error is reduced towards 0, we observe that it takes longer to accumulate measurements before making updates. } 
\label{Fig:DOC}
\end{figure}

Figure \ref{Fig:DOC} illustrates the performance of this basic DOC protocol when calibrating a $\gx$ gate in a setting where the optimal calibration value $\etaopt$ drifts according to a discrete random walk per Eq.~\eqref{Eq:RandomWalkUpdate}, with $\Delta\eta_0 = 0.15$, $\ell = 0.001$, $r = 6$, and $n=2$. We observe that as the error decreases, increasingly many samples are required to observe $n$ failures. The larger we set $r$, the faster we can update the control parameter. 

\subsection{Accelerated convergence and robustness to SPAM error}\label{sec:maintextadaptiveDOC}

Operating the DOC calibration protocol with fixed circuits can have drawbacks. With perfect state preparation and measurement, the expected number of shots required to observe $n$ failures is $n/(r^2 \alpha^2 \Delta\eta^2)$. So as gate calibration improves, the experimental overheads can become enormous. But a subtler problem emerges in the presence of errors in state preparation and measurement. Once the rate of errors due to miscalibration falls below the SPAM error rate, the protocol will be unable to reliably improve gate performance. By misdiagnosing SPAM errors as arising from gate calibration error, it will overestimate the gate error and induce overly large calibration steps in effectively random directions. The gate error will not be able to improve much beyond the point where the SPAM error is equal to the miscalibration error in the circuit. This bounds the gate error rate to approximately $\epsilon_{\rm SPAM}/r^2$. 

We can combat both effects by adaptively adjusting the circuit depth, $r$. This increases the rate of miscalibration-induced errors, which both reduces the expected number of shots required to see $n$ failures, and decreases the relative impact of SPAM errors. In App.~\ref{App:Sub:DOCHeuristic}, we consider a heuristic approach for tuning $r$ that operates by increasing the number of circuit repetitions if a threshold number of failures is not observed within a specified number of steps and decreasing $r$ if the threshold number of failures occurs too quickly.

\subsection{Generalizing to Multi-Parameter Case}\label{sec:multiparameterDOC}

For the single-parameter tuning case discussed thus far, the DOC protocol's task at each update step reduces to selecting between two choices that appear equally favorable (i.e., a positive or negative correction). In higher-dimensional control spaces, however, the ambiguity grows dramatically. The update step can no longer uniquely resolve sign ambiguities in a higher‐dimensional parameter space without further structure. This situation keeps DOC calibration from generalizing naturally to multiparameter settings. 

To see this, consider a situation where multiple control variables influence the outcome probability of a given definite-outcome circuit. This is analogous to the multiparameter situation in the IOC setting, but now the equation for the outcome probability is a quadratic, rather than linear, equation in the unknowns. Taken alone, the equation will be underdetermined. As in the IOC case, we can address this situation by defining additional circuits. This, in general, results in a set of coupled quadratic equations. We cannot solve these equations analytically anymore. Further, there will generally be a multitude of (greater than 2) solutions. It is this added ambiguity that prevents us from straightforwardly generalizing to multiparameter tuning and using the sign alternation scheme. 
As such, for the present work the IOC protocol appears to be better suited to simultaneous calibration of multi-parameter controls. 

It is, however, feasible to run multiple independent instances of DOC calibration in parallel to simultaneously tune multiple parameters. In Sec. \ref{sec:DOCqec}, we explore a path for leveraging this prospect in the context of enabling real-time drift compensation conditioned on the outcomes of quantum error correction syndrome data.

\subsection{Drift mitigation during error correction}
\label{sec:DOCqec}

Utility-scale quantum computations are expected to take hours and require high-distance quantum error correction (QEC) \cite{Caesura2025-fh,Low2025-dc, Lee2021oj, Rubin2024nh}. Over the course of the long runtime, qubit performance is expected to degrade due to drift in calibration parameters, causing an increase in the logical error rate. To maintain performance over long time scales, periodic recalibration will be necessary. Previous work \cite{fang2024caliscalpelinsitufinegrainedqubit} has considered using code deformation to remove qubits from a quantum error correcting code so that they can be periodically recalibrated. In this section, we provide a simple example demonstrating that the DOC calibration protocol can operate directly on QEC syndrome data to calibrate errors in data qubits. This premise is in line with a variety of previous works that have explored how to leverage QEC syndrome data for characterization and calibration \cite{fujiwara2014instantaneousquantumchannelestimation,fowler2014scalableextractionerrormodels,combes2014insitucharacterizationquantumdevices,florjanczyk2017insituadaptiveencodingasymmetric,huo2017learning,spitz2018adaptive,wootton2020benchmarking,PhysRevResearch.3.013292,Wagner2022paulichannelscanbe,PhysRevLett.130.200601,blumekohout2025estimatingdetectorerrormodels, PhysRevA.94.032321,kunjummen2025situcalibrationunitaryoperations,takou2025estimatingdecodingcoherenterrors,sivak2025reinforcementlearningcontrolquantum,bhardwaj2025adaptiveestimationdriftingnoise}.

During quantum error correction, cycles of mid-circuit measurements are implemented that detect the effects of errors. Decoders use these measurements to infer how these errors have affected the logical degrees of freedom. In the absence of error, syndrome measurements have definite outcomes and so provide a natural setting for realizing the DOC protocol. To explore the capability of our protocol to improve QEC performance, we've chosen a simple test case: the five qubit $[[5,1,3]]$ code \cite{PhysRevLett.77.198} in the code-capacity noise model. 

The $[[5,1,3]]$ code is a distance-3 stabilizer code defined by four stabilizer generators: $XZZXI$, $IXZZX$, $XIXZZ$, and $ZXIXZ$, with logical operators $X_L = XXXXX$ and $Z_L=ZZZZZ$, that protects one logical qubit against any of 15 possible errors on a single data qubit. It is a \textit{perfect} code, which means that each of the the 15 nontrivial error syndromes corresponds uniquely to one of the 15 nontrivial single-qubit Pauli operators. We consider exclusively coherent errors, in the code-capacity model where errors occur \textit{only} on the data qubits, between rounds of syndrome extraction (so syndrome extraction itself is flawless).  In our simulations, the 5 data qubits are initialized into the logical $\ket{0}_L$ state, and before each round of syndrome extraction the data qubits experience a unitary (coherent) noise process given by
\begin{equation}
     U(t) = \exp\left(-i \sum_{\substack{j\in{1..5} \\ k\in\{X,Y,Z\}}} \Delta\eta_k(t) \sigma_k^{(j)}\right)
     \label{eq:noisyidle}
\end{equation}
where $\sigma_X^{(j)},\sigma_Y^{(j)},\sigma_Z^{(j)}$ are the Pauli $X, Y$, and $Z$ operators acting on qubit $j$, and $\Delta\eta_k^{(j)}(t) = \eta_k^{(j)}(t) - \eta_k^{(j),\rm{opt}}(t)$ is the deviation of a control parameter from its ideal value. In our simulations the optimal values $\eta_k^{(j),\rm{opt}}(t)$ drift independently in each time step according to a random walk with step size $\ell=10^{-4}$. One time step is taken between each round of syndrome extraction, such that the number of time steps corresponds directly to the number of QEC rounds.

We use the data from repeated rounds of syndrome extraction to simultaneously run fifteen parallel DOC protocols---one to calibrate each of the three $X,Y,Z$ errors on each of the five qubits. To begin, we initialize 15 coin variables $\{c_{k}^{(j)}\}_{k=\{X,Y,Z\}}^{j=1..5}$ as +1, and 15 tuples of run counts and error counts $\{(M,m)_{k}^{(j)}\}_{k=\{X,Y,Z\}}^{j=1..5}$ as $(0,0)$. After each round, all run counts increase by one $M\to M+1$. Additionally, if we observe a non-trivial syndrome, we identify the error $E_k^{(j)}$, and increase the associated error count variable $m\gets m+1$. If at any point an error count variable reaches $n=2$, we adjust $\eta_k^{(j)} \gets \eta_k^{(j)} + c \sqrt{m/M}$, reset $(M,m)_k^{(j)} \gets (0,0)$ and flip the coin variable $c_k^{(j)} \gets - c_k^{(j)}$. 

\begin{figure*}[t] 
\centering
\includegraphics[width=2.0\columnwidth]{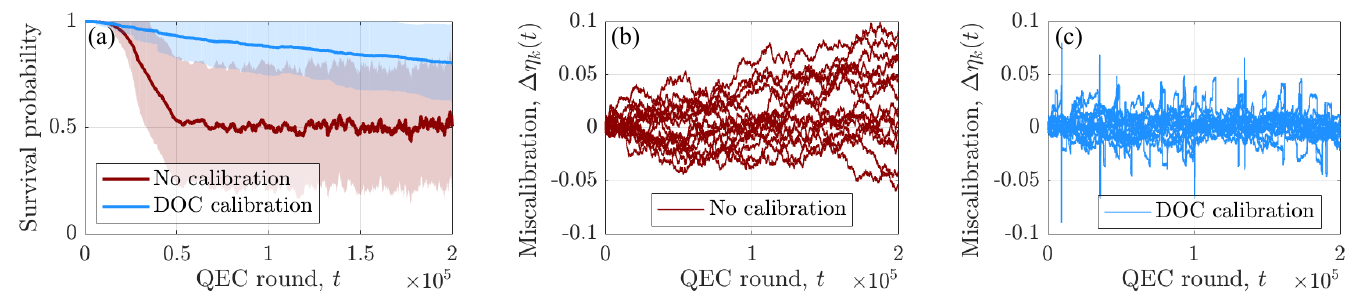} 
\caption{\textbf{Real-time DOC calibration using mid-circuit, QEC syndrome measurements in the 5-qubit code.} The performance of real-time DOC calibration, based on mid-circuit QEC syndrome measurements in the 5-qubit code, is shown. Simulations are performed using the code capacity model, with a DOC cutoff parameter of $n=2$. In panel (a) the survival probability of the state $|0\rangle_L$ is plotted against the number of rounds, $t$, of syndrome extraction for both DOC-calibrated (blue) and uncalibrated (red) cases. Solid curves show mean behavior computed over 200 realizations, while shading shows associated standard deviation. The survival probability is computed as the expectation value of the projector onto $|0\rangle_L$. In the absence of DOC calibration (red), we observe a decrease in survival probability that arises as a consequence of underlying parameter drift, described here by independent random walk processes associated with each of the 15 nominal control parameter values, as described in the text below Eq. (\ref{eq:noisyidle}). The DOC-calibrated results (blue), meanwhile, illustrate that the DOC protocol is able to sustain higher survival probabilities and more stable operation. To further illustrate the latter, panels (b) and (c) show sample trajectories of the 15 independent miscalibration errors as a function of syndrome extraction round in the absence and presence of DOC calibration, respectively. We observe that in contrast to the diverging miscalibrations in panel (b), the implementation of DOC calibration in panel (c) leads the miscalibration errors to remain approximately bounded over time, as desired.  }
\label{Fig:QEC}
\end{figure*}

Results are plotted in Fig. \ref{Fig:QEC}, which shows the accumulated survival probabilities, computed as $ \frac{\langle Z_L\rangle +1}{2}$, as a function of QEC round with and without calibration. We observe that the per-step logical error rate is significantly lower when operated with the DOC protocol. We also observe significant reductions in the number of observed error syndromes. 

Extending these results so that they represent a practical algorithm for stabilizing logical qubits in the laboratory will require moving beyond the present demonstration in at least two key aspects. First, our simulation above assumed an idealized code-capacity error model with no stochastic errors (decoherence) or readout errors.  Realistic modeling of QEC requires going to the ``noisy gate'' model -- where syndrome extraction is implemented by noisy physical operations and ancilla readout can be erroneous -- and also allowing for stochastic errors that cannot be calibrated away. Second, we have not yet considered \textit{degeneracy}, which is a general term for scenarios where syndrome data do not uniquely identify what error happened.  Perfect codes are rare, and the predominant codes in practical use (e.g. the surface code) are degenerate.  In a degenerate code, even in the code-capacity model, syndromes do not uniquely identify which qubits experienced errors. Instead they identify \textit{equivalence classes} of errors.  Degeneracy is compounded in the noisy gate model, where even for perfect codes it is not generally possible to identify which physical gate or control parameter caused a detected error.  This poses no serious problems for error correction itself, but calibration protocols need to know what specific error actually happened in order to correct the miscalibration that caused it. 

One approach to making this protocol more practical could be to intentionally amplify the error rate in part of the syndrome extraction circuit by, e.g., replacing a CNOT with a sequence of $2k+1$ CNOTs. This would have the effect of amplifying specific errors, but still allowing QEC to proceed. The calibration protocol could then operate by assuming the amplified error was the cause anytime its representative class was observed. The decoder could be made aware of the (intentionally) increased error rate.

\section{Discussion}

The IOC and DOC protocols constitute general approaches for fast feedback calibration of quantum hardware. They are robust to drift, require only minimal computational overheads, and use experimental shots very efficiently. However, realizing these benefits hinges on the capabilities of the classical control system, which can vary significantly across quantum hardware. For instance, few quantum computers today are able to update control parameters after each experimental run. In extreme cases, systems that use arbitrary waveform generators may take several seconds, or even minutes, to upload newly adjusted waveforms. In such instances, calibration protocols like ours, which rely on fast feedback, can incur significant classical communication overhead, severely limiting their performance. Conversely, some FPGA-based controllers can adapt the shape of control waveforms almost instantly in response to measured circuit outcomes. For systems that fall in between these extremes, achieving optimal performance will require a careful balance between the time allocated to data collection and the time required for updating calibration parameters. Additionally, performing calibration on multiple parameters and multiple qubits can cause rapidly scaling complexity for the classical control system. With lightweight digital implementations, these calibration protocols could be packaged into multiple calibration engines which operate in parallel on an FPGA to rapidly tune multiple parameters. In an extension to this work \cite{IOCImplementation}, we demonstrate an FPGA implementation of the IOC protocol which is capable of this sort of parallelization. This extension additionally demonstrates simulations of cryogenic CMOS ASIC implementation, and discusses how the latency between parameter updates affects IOC calibration performance.

The IOC and DOC protocols demonstrate similar convergence rates when calibrating the same $\gx$ gate in the examples given in this work, with the IOC protocol generally converging to better steady-state gate fidelity due to its ability to calibrate shot-by-shot at the cost of more strict control requirements. The primary motivation for choosing one protocol over another is centered around the circuit under calibration. In the case of this work, the $\gx$ gate with a target $\pi/2$ rotation angle naturally forms indefinite-outcome circuits by applying a single gate to an initialized $\ket{0}$. The DOC protocol is particularly well suited for circuits such as those used for syndrome extraction, where measurement data gathered during the course of syndrome extraction for error correction could itself be used for determining control parameter updates.

Like gradient descent, the protocols are robust to model uncertainty, and function reasonably well provided the system and model gradients agree in their sign. That is, IOC approaches would fail if the model implied we should \textit{increase} a parameter when it should be decreased. But if a model says we should take a step size that is twice or half as large as is optimal, the error won't significantly impact the convergence of our algorithms. This robustness comes from the self-correcting nature of the protocols, and is enhanced by a set of heuristic techniques for adaptively adjusting the depth of the test circuit and the size of the corrections. 

Although we focused our discussion in this work on the tuning of $\gx$ gates with both the IOC and DOC protocols, the protocols are meant to generalize to classes of circuits that produce indefinite or definite outcomes. The exact details of the calibration including the variety of tunable parameters, convergence rates, and some hyperparameter choices will vary based on the exact circuit and system under calibration. The examples we discussed in this work are in no way exhaustive of the space of circuits which could be calibrated by these protocols, and we expect to see exciting extensions of the IOC and DOC protocols to suit specific needs.

\subsection{Future work}

The protocols introduced here are designed to be implementation-agnostic, capable of functioning equally well across platforms. However, various quantum computing architectures can operate on vastly different timescales and are subject to distinct noise environments (for example, superconducting qubits as compared to trapped ions). While the adaptive algorithms discussed in Apps.~\ref{App:Scheduling} provide a strong foundation, there is significant potential for improving these approaches by tailoring them to system-specific details. For example, incorporating knowledge of the underlying drift structure or optimizing measurement and reset strategies to minimize latency could further enhance their performance.

Advances in understanding and modeling quantum drift \cite{proctor2020detecting,liu2023enabling} suggest that calibration protocols could benefit from leveraging system-specific drift characteristics. Systems exhibiting slow or structured drift may be particularly well-suited to hybrid strategies that combine fast feedback with predictive modeling of the underlying noise environment. Such approaches could dynamically adjust the update rate based on inferred drift patterns, optimizing calibration efficiency. For instance, Ref.~\cite{PhysRevApplied.9.064042} demonstrated that the future state of a noisy environments can be forecast from the results of earlier measurements. Incorporating these forecasts into the drift mitigation protocol could allow for significantly improved performance. 

The use of spectator qubits for real-time drift detection presents another exciting possibility for enhancing calibration protocols. Recent experiments have demonstrated the ability to monitor error syndromes via entangled or idle qubits \cite{majumder2020real,doi:10.1126/science.ade5337,PhysRevA.107.L030601,PhysRevA.107.032401,debry2025realtimemagneticfieldnoise}. Integrating IOC or DOC updates with such passive sensing mechanisms could enable background calibration to run in parallel with computation, paving the way for continuous, non-disruptive drift mitigation. This approach would allow quantum systems to maintain high performance even in the presence of dynamic noise environments.

Finally, reinforcement learning offers a compelling alternative to the heuristic-based adaptive algorithms discussed in this work. Reinforcement learning approaches have demonstrated promise in optimizing complex qubit control and calibration tasks \cite{PRXQuantum.2.040324,PhysRevApplied.18.024033, sivak2023real, reuer2023realizing, sivak2025reinforcementlearningcontrolquantum}, and system architectures \cite{caldwell2025platformarchitecturetightcoupling} compatible with running the needed real-time computations, and additionally supporting automated, low-latency control feedback, could significantly enhance calibration capabilities. 

\acknowledgements
The authors thank Christian Arenz, Kristin Beck, Fernando Calderon-Vargas, Yujin Cho, John Gamble, Matthew Grace, Erik Nielsen, Yaniv Rosen, and Mohan Sarovar for productive and enjoyable conversations. This work was supported by the Department of Energy Early Career Research Program, and the Laboratory Directed Research and Development program (Project 236997) at Sandia National Laboratories,  a multimission laboratory managed and operated by National Technology \& Engineering Solutions of Sandia, LLC, a wholly owned subsidiary of Honeywell International Inc., for the U.S. Department of Energy’s National Nuclear Security Administration under contract DE-NA0003525. SAND2025-14829O. This paper describes objective technical results and analysis. Any subjective views or opinions that might be expressed in the paper do not necessarily represent the views of the U.S. Department of Energy or the United States Government. NEM acknowledges the support of the DoD National Defense Science and Engineering Graduate (NDSEG) Fellowship program and acknowledges the funding of the Air Force Office of Scientific Research as his sponsoring agency for portions of this work. This research was funded in part by the U.S. Department of Energy, Office of Science, National Quantum Information Science Research Centers, Quantum Systems Accelerator under Air Force Contract No. FA8702-15-D-0001. Any opinions, findings, conclusions or recommendations expressed in this material are those of the author(s) and do not necessarily reflect the views of the Department of Energy.

\bibliography{bibliography}

\appendix

\section{Simulated comparison of calibration methods}
\label{App:TraditionalCal}

\begin{figure*}[t] 
\centering
\includegraphics[width=1\textwidth]{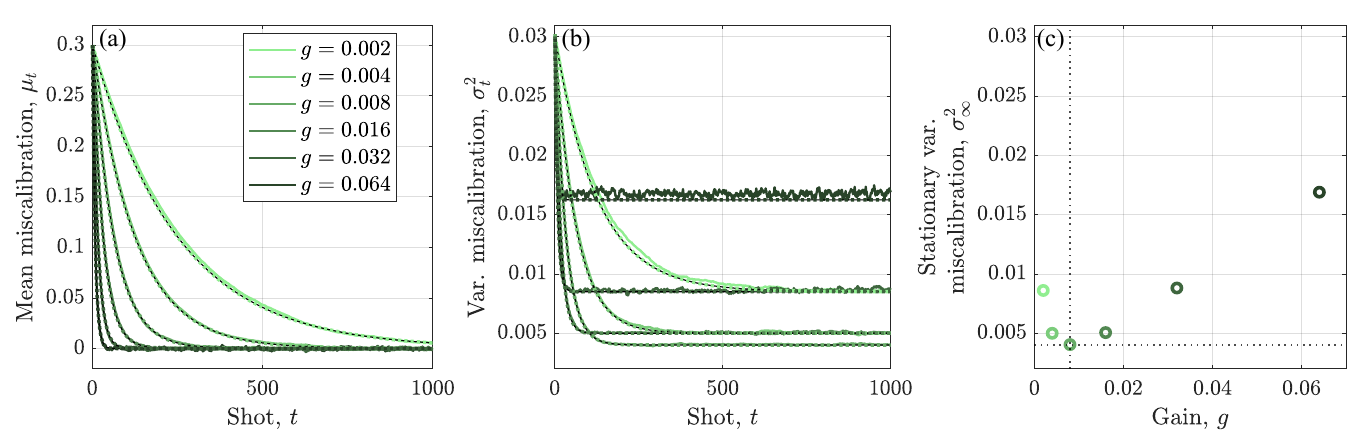} 
\caption{\textbf{Simulating dynamic and stationary behavior of the IOC protocol.} We compare the analytical results from Sec. \ref{App:Sub:IOCBehavior} with numerical simulation for executions of the IOC protocol with different choices of gain, $g$. Panels (a) and (b) present results analogous to those presented without drift in the main text in Fig. \ref{Fig:analyticalcomparisonmaintext}; here, we incorporate drift and compare our numerical findings to analytical predictions. In panels (a) and (b), solid curves show the results of numerical simulations, where the mean is taken over 10,000 realizations of the IOC protocol with $r=1$ and drift described by a random walk with $\ell = 0.008$. Different colors correspond to different values of the gain, $g$. In (a) and (b), the dotted curves of like colors (connected with thin black solid curves for visibility) are computed from the analytical expressions in Eq. (\ref{eq:mudiffeqsoln}) and (\ref{eq:vardiffeqsoln}), respectively. Panel (a) presents the convergence of the mean, $\mu_t$, of the parameter deviation, $\Delta\eta_t$, as a function of the step, $t$, when the initial offset is set to $\Delta\eta_0 = 0.3$ for all trajectories. Panel (b) shows the convergence of the variance, $\sigma^2_t$ of the parameter deviation as a function of $t$ for initial offsets sampled uniformly randomly from $\Delta\eta_0\in [-0.3, 0.3]$, such that $\mu_0 = 0$ in line with the assumption behind Eq. (\ref{eq:vardiffeqsoln}). In both (a) and (b), we observe good agreement between theory and simulation. Panel (c) shows the stationary variance, $\sigma^2_\infty$, of the parameter deviation as a function of the gain, $g$. Point markers correspond to the numerically computed stationary variances from the simulation results in panel (b). Thin dotted black lines mark the optimal stationary variance, per Eq. (\ref{eq:statvar}), and the associated value of the gain, per Eq. (\ref{eq:solutiontovardynamics}).}
\label{Fig:CompareIOCdynamicstotheory}
\end{figure*}

In this appendix, we provide further details regarding the comparison presented in Fig. \ref{Fig:infidvsdutycycle} of the main text, where we compare the performance of the IOC and DOC protocols for calibrating a $\gx$ gate against that of a Rabi calibration protocol. The latter is meant to represent the type of calibration protocol used in many hardware calibration experiments today.
In our comparison, we observe that, for fixed gate infidelities, there can be a benefit of the IOC and DOC calibration protocols over Rabi calibration when it comes to total time the quantum system must spend performing calibration compared to being able to perform useful experimental circuits. We call this the calibration duty cycle, which we compute as
\begin{equation}
    D=\frac{T_c}{T_c+T_e},
\end{equation} 
with $T_c$ the number of shots spent performing calibration and $T_e$ the number of shots spent performing other circuits (or, in the case of our simulations, allowing drift to propagate in the model without performing calibration). For the IOC protocol, for example, $D=1\%$ corresponds to performing $1$ calibration shot, then waiting for drift to propagate for (the equivalent of) $99$ shots, noting that $T_{c,IOC}$ is always equal to $1$ unless we use batched approaches described in \ref{App:Sub:IOCBatched}. We observe that the IOC and DOC protocols maintain low errors even when spending a smaller fraction of time calibrating than batched protocols such as the Rabi-like protocol.

The protocol for the batched Rabi calibration proceeds as follows:

\begin{enumerate}
    \item For $r_i$ in some range $[0,r-1]$ of circuit repetitions, estimate the output probabilities of $\gx^{r_i}$ using $N_{batch}$ shots per circuit (totaling $N_{batch}\cdot r$ shots). Include drift, SPAM and per-gate depolarization.
    \item Fit the output probability curve to:
        \begin{equation}
            P(r_i) = a \cdot (b^{r_i}) \cdot (\sin(\theta_{est}\cdot r_i/2)^2) + c
        \end{equation}
        where $a$ estimates the SPAM error as $p_{SPAM}=1-a$, $b$ estimates the per-gate depolarization as $p=1-b$, $c$ is an estimated offset, and $\theta_{est}$ is the estimated gate rotation angle which is targeted to be $\pi/2$. We also seed the fitter with some knowledge of reasonable parameter ranges. Here, we expect $0.9\leq a \leq 1$, $0.9 \leq b \leq 1$, and $\pi/4 \leq \theta \leq 3\pi/4$.
    \item From the estimate of $a$, $b$, $c$, and $\theta_{est}$ using the least-squares-based \texttt{curve\_fitter} function from \texttt{scipy.optimize}, correct the estimated gate error $\hat\delta_t=\pi/2-\theta_{est}$ by applying $\eta_{t+1}=\eta_t+\hat\delta_t/\alpha$.
    \item Allow drift to occur in the gate model for $T_{e,Rabi} = N_{batch}\cdot r\cdot (\frac{1}{D} - 1)$ rounded to an integer number of shots, then repeat the calibration process.
\end{enumerate}

For the implementation in Fig. \ref{Fig:infidvsdutycycle}, we use hyperparameters of $r=20$ and $N_{batch}=20$ shots per circuit.

For the IOC protocol, we set the circuit repetitions to a constant $r=13$, which we experimentally determined as the highest $r$ we could use without falling into the wrong optimization well, and set the gain to $g=\sqrt{T_{e,IOC}+1}\cdot \ell\cdot r$ for $r=13$, its optimal value given a waiting period between calibration shots. Similarly for the DOC protocol, we set the circuit repetitions to a constant $r=10$.

For all three protocols, we model random walk drift in $\eta_{opt}$ with $\ell=0.001$ occurring at each shot, per-gate depolarization with $p=0.001$ and SPAM error with $p_{SPAM}=0.01$, and $\Delta\eta_0=0$. We run each of the protocols for $100$ trajectories of $100,000$ shots each. At each shot, we calculate the infidelity of the tuned $\gx$ gate. 

To produce the plot in Fig. \ref{Fig:infidvsdutycycle}(c), we specifically calculate the mean process infidelity over the course of the full experiment length of $100,000$ shots. The plot then shows the median and interquartile range of these experiment means at each duty cycle over $100$ calibration trajectories. The process infidelity is computed as the entanglement infidelity \cite{PRXQuantum.6.030202}:
\begin{equation} \label{eq:entanglementinfidelity}
    \mathcal{I} = 1-\frac{1}{d^2}   
        \mathsf{Tr}\left[ 
             \Lambda_G\Lambda_U^{-1} 
            \right],
\end{equation} with $d=2$ and $\Lambda_G$ and $\Lambda_U$ denoting the Pauli transfer matrix representations of the miscalibrated $\ux(\delta_t)$ gate, with added depolarization error, and the target $\ux(0)$ gate, respectively. Equation (\ref{eq:entanglementinfidelity}) reduces in the special case of two unitary operations to Eq. (\ref{eq:unitaryinfidelity}) in the main text.

\section{IOC protocol notes}
\label{App:IOC}

In this appendix, we discuss additional theoretical details and simulation analyses relevant to the IOC protocol. In Sec. \ref{App:Sub:IOCBehavior} we lay out the theory associated with our discussion in sections \ref{sec:IOCsimpleexample} and \ref{Sec:LowLatencyDriftMitigation} of the main text. We go on in Sec.  
\ref{App:Sub:GradientDescent} by outlining the relationship between IOC calibration and stochastic gradient descent. This is followed by an analysis in Sec. \ref{App:Sub:IOCBatched} comparing the single-shot IOC protocol with batched, multi-shot protocols, and a discussion of the regime where the former outperforms the latter.

\subsection{Dynamic and stationary behavior}
\label{App:Sub:IOCBehavior}

Here, we provide additional theoretical details for the IOC protocol. We focus this discussion on the simple, motivating example of calibrating a $\gx$ gate, as introduced in Sec. \ref{sec:IOCsimpleexample} of the main text, where $s$ is a sensitivity parameter and $\Delta\eta = \eta-\etaopt$ describes the deviation between the tunable parameter $\eta$ and its optimal value $\etaopt$. Recall from the main text:
\begin{equation}\label{eq:app:etalinear}
    P(z_t=\pm 1|\Delta\eta_t) \approx 1/2 \mp s\Delta\eta_t,
\end{equation}
i.e., that the application of $\gx$ to the state $|0\rangle$ leads to outcome probabilities with an approximately linear relationship to $\Delta \eta$. The IOC protocol then aims to minimize $\Delta\eta$ over a sequence of steps indexed by $t$. At each step, $\eta_t$ is updated according to Eq. (\ref{eq:IOCupdaterule}) in the main text,
\begin{equation}\label{eq:app:updaterule}
    \eta_{t+1} = \eta_t+ \gain z_t/s,
\end{equation}
where $z_t\in\{-1,1\}$ is the outcome of measuring $\sigma_z$ at step $t$, and $\gain$ is a gain parameter. In alignment with Eq. (\ref{Eq:RandomWalkUpdate}) of the main text, we consider the situation that $\etaopt[t]$ drifts according to a discrete random walk, such that
\begin{equation}\label{eq:app:rwupdate}
\etaopt[t] = \etaopt[t-1]+q_{t-1}\ell,
\end{equation}
where $\ell$ is the magnitude of each random walk step and $q_t\in\{-1,1\}$ dictates the sign of the random walk step. We assume both signs are equiprobable for all steps $t$. 

In the following, we provide additional details regarding the theoretical discussion in the main text surrounding the dynamic and stationary behavior of the IOC protocol, as well as gain selection (and scheduling). For the latter, the aim is to select $\gain$ in a manner that gives desirable convergence behavior, as captured by the mean and the variance of $\Delta\eta_t $ produced by the IOC protocol. We denote these quantities at step $t$ by $\mu_t\equiv \E[\Delta\eta_t]$ and $\sigma^2_t \equiv\E[\Delta\eta_t^2]-\mu_t^2$, respectively. 

We begin by considering the dynamics of $\mu_t$. Combining Eqs. (\ref{eq:app:updaterule}) and (\ref{eq:app:rwupdate}) gives
\begin{equation}
    \Delta\eta_t = \Delta\eta_{t-1}-q_{t-1}\ell +z_{t-1}\frac{\gain}{s}.
\end{equation}
Taking the expectation leads to
\begin{equation}\label{eq:app:intermediatestepiniocderivation}
\begin{aligned}
    \mu_t= \mu_{t-1}-\E[q_{t-1}]\ell +\E[z_{t-1}]\frac{\gain}{s}.
\end{aligned}
\end{equation}
For an unbiased random walk, $\E[q_{t-1}] = 0$. Then, from Eq. (\ref{eq:app:etalinear}), we obtain 
\begin{equation}
    \begin{aligned}
    \E[z_t] &= \sum_{z_t =\pm 1}  P(z_t)z_t =\int d\Delta\eta \sum_{z_t = \pm 1} P(z_t|\Delta\eta_t) P(\Delta\eta_t)z_t \\
    &=-2s\E[\Delta\eta_t],
    \label{eq:E(z)}
    \end{aligned}
\end{equation}
which can be substituted into Eq. (\ref{eq:app:intermediatestepiniocderivation}) to give the difference equation 
\begin{equation}
\begin{aligned}
    \mu_t &= \mu_{t-1}-2\gain\mu_{t-1}
\end{aligned}
\label{eq:mu}
\end{equation}
describing the dynamics of $\mu_t$, whose solution at step $t$ is given by
\begin{equation}
    \mu_t = (1-2\gain)^t \mu_{0},
\end{equation}
where $\mu_0$ represents the initial condition, which may in general be biased away from $\mu_0 = 0$. We observe that for $t\rightarrow\infty $ we obtain $\mu_\infty = 0$, and that the rate of convergence can be increased by increasing $\gain\in[0,1/2)$. 

When $g$ is small, the dynamics generated by the difference equation in Eq. (\ref{eq:mu}) can be well approximated by the differential equation $\frac{d\mu(\tau)}{d\tau} = -2\gain\mu(\tau)$, whose solution is given by 
\begin{equation}
    \mu(\tau) = \mu(0) e^{-2\gain\tau}
    \label{eq:mudiffeqsoln}
\end{equation}
for $d\tau = 1$. This is validated numerically in Fig. \ref{Fig:CompareIOCdynamicstotheory}(a).

We now consider the dynamics of $\sigma^2_t\equiv\E[\Delta\eta_t^2]-\mu_t^2$, following the same procedure to evaluate $\E[\Delta\eta_t^2]$ as used above to evaluate $\E[\Delta\eta_t]$. Taking $\E[q_t] = 0$ and $\E[z_t^2] = \E[q_t^2] = 1$, we obtain
\begin{equation}
\E[\Delta \eta_t^2] =\E[\Delta \eta_{t-1}^2] + \frac{\gain^2}{s^2} + \ell^2 + 2\E[\Delta \eta_{t-1} z_{t-1}]\frac{\gain}{s}.
\end{equation}
We evaluate the final term on the right-hand side as
\begin{equation}
    \begin{aligned}
    \E[\Delta\eta_tz_t] &=\int d\Delta\eta_t\,dz_t\, P(z_t|\Delta\eta_t) P(\Delta\eta_t)\Delta\eta_tz_t \\
    &=-2s\E[\Delta\eta_t^2].
    \end{aligned}
\end{equation}
Putting everything together then yields the difference equation for the variance,
\begin{equation}
\begin{aligned}
    \sigma_t^2 &=\sigma_{t-1}^2 + \gain^2 / s^2 + \ell^2 - 4\gain\sigma^2_{t-1} -4\gain^2\mu_{t-1}^2,
    \label{eq:variancedifference}
\end{aligned}
\end{equation}
whose solution is given by
\begin{equation}
\begin{aligned}
    \sigma^2_t &= \left(\sigma^2_0 - \frac{\gain}{4s^2}-\frac{\ell^2}{4\gain}\right)(1-4\gain)^t+\frac{\gain}{4s^2}+\frac{\ell^2}{4\gain}\\
    &\quad -4\gain^2(1-4\gain)^{t-1}\mu_0^2.
    \end{aligned}
    \label{eq:vardiffeqsoln}
\end{equation}

For small $g$, the dynamics generated by the difference equation in Eq. (\ref{eq:variancedifference}) can be well approximated by the differential equation $\frac{d\sigma^2(\tau)}{d\tau} = \gain^2 / s^2 + \ell^2 - 4\gain\sigma^2(\tau)-4k^2\mu^2(\tau)$, whose solution for $\mu(\tau) = 0,\,\forall \tau$ is given by 
\begin{equation}
    \sigma^2(\tau) = \left( \sigma^2(0) - \frac{k}{4s^2}-\frac{\ell^2}{4\gain}\right)e^{-4\gain\tau}+\frac{\gain}{4s^2}+\frac{\ell^2}{4\gain},
    \label{eq:solutiontovardynamics}
\end{equation}
which is validated numerically in Fig. \ref{Fig:CompareIOCdynamicstotheory}(b) and (c).

Returning to the prospect of gain selection, we can find the value of $\gain$ that minimizes the stationary variance $\sigma^2_\infty$, given by
\begin{equation}
\sigma^2_\infty = \frac{\gain}{4s^2}+\frac{\ell^2}{4\gain}, 
\label{eq:statvar}
\end{equation}
by solving $\frac{\partial\sigma^2_\infty}{\partial \gain}=\frac{1}{4s^2}-\frac{\ell^2}{4\gain^2}=0$ for $\gain$, yielding
\begin{equation}
   \gain_\infty=\ell s.
    \label{eq:optkinfty}
\end{equation}
In practice, $\ell$ may not be known.

\begin{figure}[t] 
\centering
\includegraphics[width=\columnwidth]{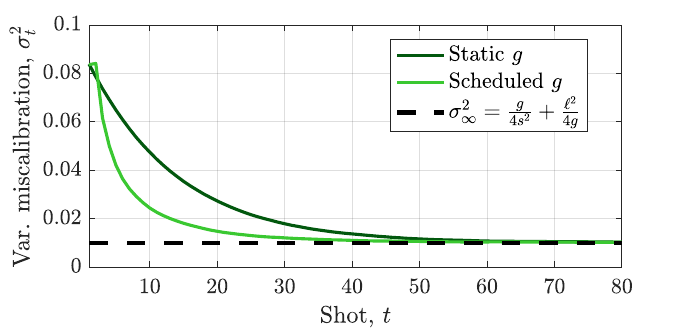} 
\caption{\textbf{Calibrating with the analytically-derived gain schedule.} We compare the performance of the IOC protocol using the gain schedule in Eq. (\ref{eq:kschedule}) against the performance using a static gain. To evaluate Eq. (\ref{eq:kschedule}), 50,000 realizations of the IOC protocol with $\ell = 0.02$, $r=1$, and $\Delta\eta_0$ sampled uniformly randomly between $-0.5$ and $0.5$, are utilized to compute the mean $\mu_t$ and variance $\sigma^2_t$ of the parameter deviation $\Delta\eta_t$ at all steps $t$. For the scheduled gain case, the gain is initialized using $\gain_0 = 0$, and then updated using the schedule in Eq. (\ref{eq:kschedule}) to determine $\gain_t$ for $t>0$. The initial increase in the variance associated with the scheduled case is due to the lack of calibration action when $g=0$. For the static gain case, $\gain_t = \ell s$ is utilized at all steps $t$ per Eq. (\ref{eq:optkinfty}), as this is the value of $\gain$ that should produce the lowest stationary variance per Eq. (\ref{eq:optkinfty}). We observe that both the static (dark green) and the scheduled (light green) gain lead to a variance of the miscalibration that converges asymptotically to the minimal stationary variance predicted in Eq. (\ref{eq:statvar}). Gain scheduling according to Eq. (\ref{eq:kschedule}) leads to faster transient convergence of $\sigma^2_t$. We note that the scheduled gain converges to $g=\ell s$ automatically, without knowledge of $\ell$.}
\label{Fig:gainschedulingnoapprox}
\end{figure}

We can also allow the gain to be $t$-dependent, such that we have a gain schedule. From the results thus far, an intuitive choice is to initialize $\gain$ to be large, in order to accelerate the convergence of $\mu_t$, and then shrink $\gain$ towards $\gain_\infty = \ell s$ over time in order to minimize the stationary variance. Concretely, we can seek a gain schedule to maximize the rate of change of ${\sigma}^2_t$ so that it converges rapidly to $\sigma^2_\infty$. Evaluating $\frac{\partial (\sigma^2_t - \sigma^2_{t-1})}{\partial \gain}=\frac{2\gain}{s^2}-4\sigma^2_t-8\gain\mu^2_t=0$ from Eq. (\ref{eq:variancedifference}) and solving for $\gain$ yields the gain schedule
\begin{equation}
    \gain_t = \frac{2\sigma^2_t s^2}{1-4s^2\mu^2_t},
    \label{eq:kschedule}
\end{equation}
noting that taking the limit as $t\rightarrow \infty$ returns $\gain_\infty = \ell s$ as expected. This gain schedule has been validated in numerical simulation, with results plotted in Fig. \ref{Fig:gainschedulingnoapprox}. In practice, Eq. (\ref{eq:kschedule}) can rarely be evaluated directly due to its dependence on $\mu_t$ and $\sigma^2_t$, which are typically not known or observable quantities. We discuss a more practical implementation of this gain scheduling protocol in the approximate error estimation approach described in App. \ref{App:SubSub:ApproximateErrorEstimation}, where $\mu_t$ and $\sigma^2_t$ are replaced by estimates computed from the measurement record. 

\subsection{Relationship to stochastic gradient descent}
\label{App:Sub:GradientDescent}

The IOC protocol can be viewed as an extreme example of stochastic gradient descent (SGD) \cite{SGD} which is an iterative, gradient-based optimization algorithm widely used in machine learning. Stochastic gradient descent aims to minimize a loss function that can be represented as a sum of per-sample loss functions: $\mathcal{L}(\bm{\eta}) = \frac{1}{N} \sum_{i=1}^N \mathcal{L}_i(\bm{\eta})$. Standard gradient descent computes the gradient and updates the parameters $\bm{\eta}$ in the direction that reduces the total loss. In contrast, SGD estimates the gradient using only a subset of the per-sample losses. In machine learning applications, computing a full gradient is often impractical due to the large size of the training data set. Instead, the data is typically divided into mini-batches, with gradient estimates computed via backpropagation. Because each mini-batch is a different subset of the data, these gradient estimates are inherently noisy. However, under standard assumptions, such as bounded variance of the stochastic gradients and an appropriately diminishing learning rate, convergence guarantees for SGD can be established.

The IOC protocol can be considered as optimizing a simple quadratic cost functional: 
\begin{equation}
    \mathcal{L}(\bm{\eta}) = \frac{1}{2} \vert \Delta\bm{\eta} \vert^2. 
\end{equation}
The gradient of this function is given by $\nabla_{\bm{\eta}} \mathcal{L}(\bm{\eta}) = \Delta\bm\eta$. 

The miscalibration $\Delta\bm{\eta}$, and thus the gradient of the loss function, can be inferred from linear tomographic measurements on an ensemble of circuits $\{C^{(k)}\}_k$ as follows. In the large data limit, the observed frequency of measuring outcome $z$ after running circuit $C^{(k)}$ will converge to the true probability. These probabilities can be approximated to first order as:
\begin{equation}
    p_z^{(k)}\left(\Delta\bm\eta\right) = p_{z}^{(k)}(\bm\eta_{\rm{opt}}) + \bm{s}_z^{(k)} \cdot \Delta\bm{\eta}
\end{equation}
This is a linear family of equations that can be inverted to solve for $\Delta\bm\eta$ from the observed probabilities. To do so, first define $\bm{p}(\Delta\bm{\eta})$ as the vector of the outcome probabilities 
\begin{equation}
\label{eq:jacobian}
    \bm{p}(\bm{\eta}) = \left(p^{(1)}_0(\bm{\eta}), \cdots, p^{(1)}_{2^n-1}(\bm{\eta}), \,
                        p^{(2)}_0(\bm{\eta}), 
                        \cdots \right)^{\mathsf{T}},
\end{equation}
and so $\Delta\bm{p}(\bm{\eta}) = \bm{p}(\bm{\eta}) - \bm{p}(\bm{\eta}_{\rm{opt}})$. In terms of the Jacobian matrix $\mathcal{J}$, defined in Eq.~\eqref{eq:jacobian}, we have
\begin{equation}
\label{eq:axeqb}
    \Delta\bm{p} = \mathcal{J}\cdot\Delta\bm{\eta}
\end{equation}
In general, $\mathcal{J}$ is rectangular, so the least-squares solution to Eq.~\eqref{eq:axeqb} can be found using  Moore-Penrose pseudoinverse:
\begin{equation}
    \Delta\bm{\eta} = \mathcal{J}^+ \cdot \Delta\bm{p}
\end{equation}

In the absence of error, indefinite-outcome circuits lead to measurement probability distributions that are uniform over a subset of possible outcomes. In most cases, calibration errors cannot move population out of the support of this distribution at first order.

Conservation of probability guarantees that the columns of the Jacobian sum to zero, so it follows that $\mathcal{J}^+\cdot \bm{p}(\bm{\eta}_{\rm{opt}}) = 0 $ and we can replace the probability error with the probability:
\begin{equation}
    \Delta\bm{\eta} = \mathcal{J}^+ \cdot \bm{p}
\end{equation}
Finally, if we have chosen a set of circuits that are approximately uniformly sensitive to errors in all parameters, then the Jacobian will be well conditioned, i.e., its singular values will be all relatively close to one another. In this case, the pseudoinverse will be proportional to the transpose. This follows directly from the singular value decomposition of the (real-valued) Jacobian, $\mathcal{J} = U\Sigma V^\mathsf{T}$, with $\Sigma$ the diagonal matrix of singular values and $U$ and $V$ orthogonal matrices. The pseudoinverse is $\mathcal{J}^+ = V \Sigma^+ U^\mathsf{T}$, where $\Sigma^+$ is formed by taking the reciprocal of the non-zero singular values. If the singular values are all approximately the same, then $\Sigma^+ \approx \Sigma/\abs{\abs{\Sigma}}_2^2 $. Here $\abs{\abs{\cdot}}_2$ is the spectral norm, which is equal to the largest singular value. Consequently, $\mathcal{J}^+ \approx V \Sigma U^\mathsf{T} / \abs{\abs{\Sigma}}_2^2 = \mathcal{J}^\mathsf{T} / \abs{\abs{\Sigma}}_2^2$. So the relationship between the parameter error and the outcome probabilities is:
\begin{equation}
    \Delta\bm\eta \propto \mathcal{J}^\mathsf{T} \cdot \bm{p}
\end{equation}

This equation allows us to estimate the error in the calibration parameters, and thus the gradient of the loss function, in terms of the empirical distribution of the measurement outcomes. One could estimate $\mathbf{p}$ using many shots of each circuit, but in the extreme stochastic limit, we take only a \textit{single} shot of one circuit. In this case the maximum likelihood estimate of $p$ is simply the unit vector indicating which result was observed. The resulting gradient estimate is then simply the row of the Jacobian that corresponds to the observed outcome. This is the direction chosen by the IOC protocol.

\subsection{The IOC protocol without single-shot updates}
\label{App:Sub:IOCBatched}

\begin{figure}[t] 
\centering
\includegraphics[width=\columnwidth]{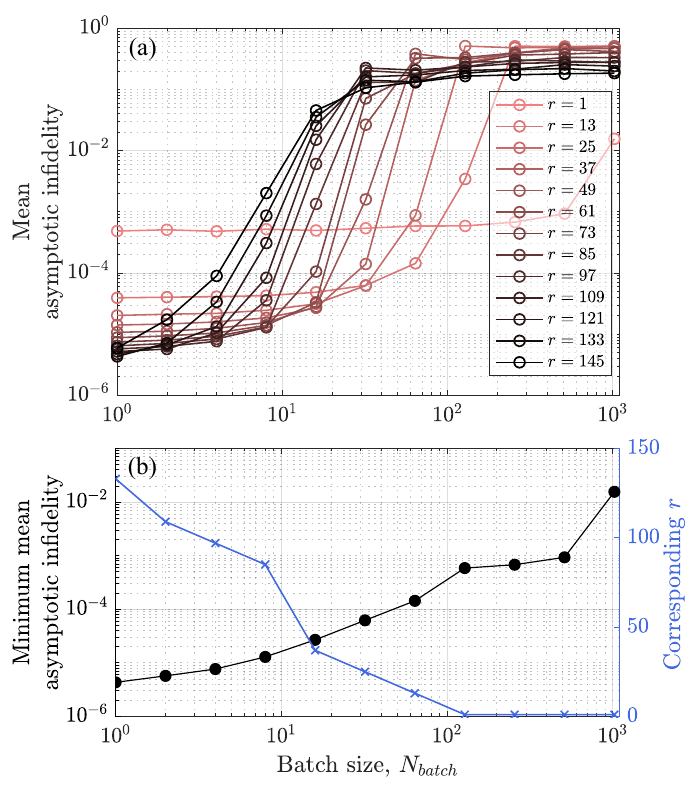} 
\caption{\textbf{Analysis of multi-shot IOC protocols.} Panel (a) plots the mean asymptotic infidelity of a $\gx$ gate as a function of batch size, $N_{batch}$, for an IOC protocol across different values of coherent gate repetitions per shot, $r$. Panel (b) presents the minimum mean asymptotic infidelity obtained as a function of batch size, where the minimum is taken over $r$, on the left $y$-axis and the corresponding $r$ values on the right $y$-axis. Here, batch size refers to the number of shots used to inform each control parameter update. The control parameter is subject to drift described by a discrete, unbiased random walk with step size $\ell=0.001$.
For all simulations, the initial miscalibration $\Delta\eta_0 = 0$ and $g = \ell s$. The curves of different colors in panel (a) correspond to different values of $r$. We see that for small $N_{batch}$, increasing $r$ allows for more precise tuning and leads to reductions in the mean infidelity, up to some threshold value (here, $\approx r=133$). Past this threshold, continuing to increase $r$ leads to increases in the infidelity. We also observe that across $r$ values, the mean infidelity decreases as we move towards the single-shot limit $N_{batch}=1$. This finding supports the investigation of single-shot IOC calibration and drift mitigation protocols in this work. }
\label{Fig:batchanalysissummary}
\end{figure}

Here we investigate how the single-shot IOC protocol compares with a batched, multi-shot version of the IOC protocol calibration protocol where $N_{batch}>1$ samples are utilized at each step to inform updates to $\eta$. We consider the calibration of a $\gx$ gate as introduced in Sec. \ref{sec:IOCsimpleexample}, modified to consider a batched update rule given by
\begin{equation}
\begin{aligned}
    \eta_{t+N_{batch}} &= \eta_{t} + \frac{1}{N_{batch}}\sum_{j=1}^{N_{batch}} z_{t_j} \gain /s
    \label{eq:updaterule_batched}
    \end{aligned}
\end{equation}
where $t$ indexes the step of the IOC protocol, but now each step contains $N_{batch}$ shots, indexed by $j$ such that the $j$-th shot of the $t$-th step is indexed by $t_j$ and corresponds to the point where we have taken $tN_{batch}+j$ total shots. We observe that for $N_{batch}=1$ we recover the basic IOC protocol.

Figure \ref{Fig:batchanalysissummary}(a) illustrates how the mean asymptotic infidelity of a drifting $\gx$ gate, stabilized by the batched IOC protocol, varies as a function of the batch size, $N_{batch}$. Infidelity calculations use the unitary formula in Eq. (\ref{eq:unitaryinfidelity}) with $W = \ux (\delta_t)$ and $U = \ux(0)$. The mean is computed over the final 10\% of shots from 500 realizations of the IOC protocol with 10240 total shots each (subdivided into batches of size $N_{batch}$). The control parameter $\eta$ is assumed to initially be on target such that $\eta=\eta_{opt}$ at $t=0$. After, $\eta_{opt}$ is subject to time-dependent drift that can be described by a discrete, unbiased random walk per Eq. (\ref{Eq:RandomWalkUpdate}) with $\ell=0.001$ that advancess between circuit executions.
For all cases in this analysis, the value of the IOC protocol gain parameter, $\gain$, is selected as $\gain=\ell s$, motivated by the discussion in Sec. \ref{Sec:LowLatencyDriftMitigation}. 

In Fig. \ref{Fig:batchanalysissummary}(a), the curves of different colors correspond to different values of $r$ used in the analysis, which denotes the number of coherent repetitions of the $\gx$ gate per shot. We observe that for mean asymptotic infidelities greater than the minimum, there appears to be a set of multiple $N_{batch}$ values that achieve similar mean asymptotic infidelities, indicating that here, there may be no benefit to doing single-shot versus multi-shot calibration protocols. However, both the mean asymptotic infidelity, as well as the variance of the asymptotic infidelity (not shown), decrease as we move towards the single-shot limit $N_{batch}=1$. In this limit, single-shot tuning becomes the most desirable option for minimizing the mean and variance of the asymptotic infidelity. This trend is displayed in Fig. \ref{Fig:batchanalysissummary}(b), where the results plotted using the left $y$-axis show the minimal value of the mean asymptotic fidelity plotted above in Fig. \ref{Fig:batchanalysissummary}(a) as a function of $N_{batch}$, where the minimum is taken over $r$. We see that the minimum achieved mean asymptotic infidelity decreases monotonically as a function of batch size, reaching its minimal value for $N_{batch}=1$ (the single-shot limit). The right $y$-axis of Fig. \ref{Fig:batchanalysissummary}(b) plots the corresponding $r$ value that minimizes the mean asymptotic infidelity.

\section{DOC protocol notes: choosing an estimator}
\label{App:DOC}

In the DOC protocol, we repeat a definite-outcome circuit and record the corresponding measurement outcomes until $n$ undesired outcomes occur (we refer to $n$ as the cutoff parameter in Sec. \ref{sec:DOC}). We can model the distribution of measurement outcomes as a negative binomial distribution with a parameter $p$, the probability of sampling undesired outcomes. As such, we can use the number of undesired outcomes, $n$, and the number of shots where desired outcomes occur while waiting for $n$ to be achieved, $k$, to form an estimate of the parameter, $\hat{p}$. 

Here, we consider two methods for estimating $\hat{p}$. First, the minimum-variance unbiased estimator (MVUE):

\begin{equation}
    \hat{p}_{MVUE}=\frac{n-1}{n+k-1}
    \label{Eq:MVUE}
\end{equation}

\noindent and second, the maximum likelihood estimator (MLE):

\begin{equation}
    \hat{p}_{MLE}=\frac{n}{n+k}
    \label{Eq:MLE}
\end{equation}

\begin{figure}
\centering
\includegraphics[width=\columnwidth]{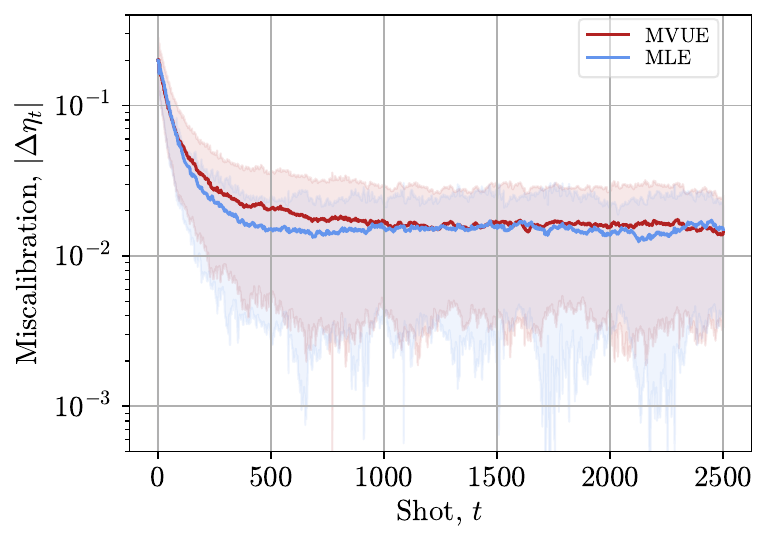} 
\caption{\textbf{Comparing the MVUE and MLE estimators for the DOC protocol.} We simulate 200 trajectories of DOC calibration using the MVUE and MLE estimates of the error probability under conditions of an initial $\Delta\eta_0=0.2$, a random walk drift with $\ell=0.001$, depolarizing SPAM error with $p_{SPAM}=0.01$, and a per-gate depolarization with $p=0.001$. We utilize a constant number of circuit repetitions $r=10$. Here, we observe that using the MLE leads to slightly faster convergence and lower stationary variance. This trend is more pronounced in the absence of drift, SPAM error, and per-gate depolarization (not pictured).}
\label{Fig:comparingestimators}
\end{figure}

Using both estimators, we simulate 200 trajectories of the DOC calibration protocol using a cutoff parameter of $n=2$ under the conditions of random walk parameter drift, depolarizing SPAM, and per-gate depolarization. These results are plotted in Fig. \ref{Fig:comparingestimators}. We find that using the MLE leads to slightly faster convergence and lower stationary variance than using the MVUE, so we choose to use the MLE estimator in all of the demonstrations in this work.

\section{Gain and circuit repetition scheduling}
\label{App:Scheduling}

In this appendix, we provide additional details about heuristic approaches we propose for adaptively adjusting the hyperparameter(s) of the IOC and DOC protocols to improve their performance. First, in Sec. \ref{App:Sub:IOCHeuristics} we discuss the two heuristic approaches introduced in Sec. \ref{Sec:IOCAdaptiveStrategies} of the main text for scheduling the gain, $g$, and the number of circuit repetitions, $r$, to improve the performance of the IOC protocol. Then, we discuss repetition scheduling for the DOC protocol in Sec \ref{App:Sub:DOCHeuristic}. 

\subsection{IOC protocol scheduling heuristics}
\label{App:Sub:IOCHeuristics}

\subsubsection{Approximate error estimation}
\label{App:SubSub:ApproximateErrorEstimation}

\begin{figure}
\centering
\includegraphics[width=0.9\columnwidth]{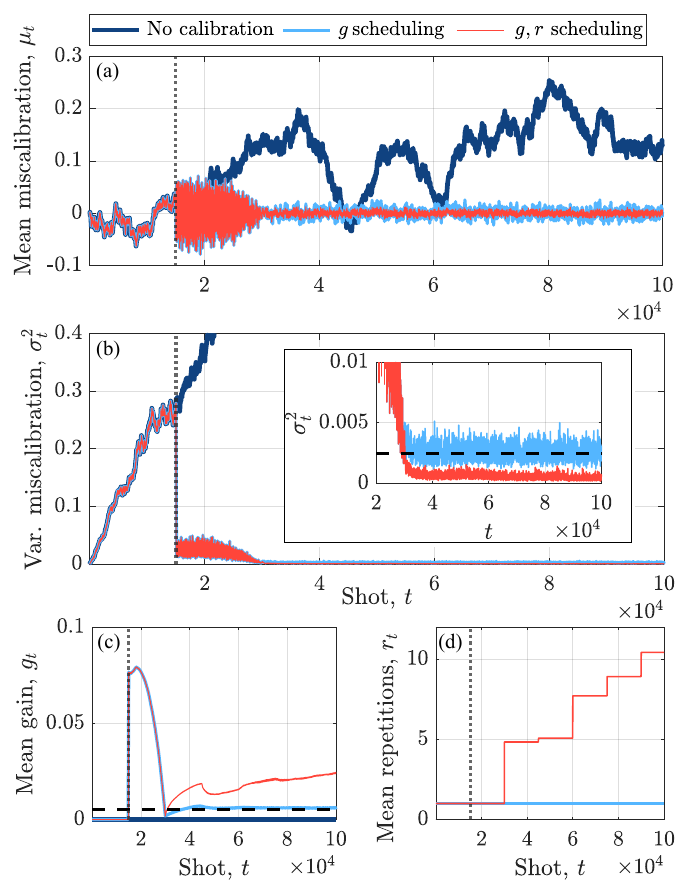} 
\caption{\textbf{Scheduling the IOC protocol with approximate error estimation.} The performance of the IOC protocol is shown for three cases: an uncalibrated baseline (dark blue), $\gain$ scheduling (light blue), and $g,r$ scheduling (red). For $g$ scheduling, the gain is adjusted using the approximate error estimation protocol in Eq. (\ref{eq:kscheduleapproximate}) with $M = 75$, $N = 200$, and a sliding window to estimate the mean and variance according to Eqs. (\ref{Eq:tildemu}) and (\ref{Eq:tildesigma}), respectively. For $g,r$ scheduling, we additionally schedule $r$ per Eq. (\ref{eq:repetitionscheduling}), with $\delta_{max}=\pi/4$, $\kappa=3$, and without a sliding window. Results are computed from 50 realizations with $\ell = 0.005$, $r_0=1$, and $\Delta\eta_0=0$. In all panels, the vertical dotted black lines mark $t=MN$ when the scheduling protocols begin. Panel (a) shows the mean, $\mu_t$, of the parameter deviation, $\Delta\eta_t$, as a function of $t$. We observe that both the $\gain$ and $\gain,r$ scheduled IOC protocols compensate for drift and stabilize $\mu_t$; the stabilization is tightest for $\gain,r$ scheduling. Panel (b) shows the variance, $\sigma^2_t$, of $\Delta\eta_t$. We observe a reduction and stabilization of the variance to very small values. The inset shows an enlarged plot of the variance, together with a horizontal dashed black line that marks the minimal stationary variance, $\sigma^2_\infty = \frac{\gain}{4s^2}+\frac{\ell^2}{4\gain}$, when $s = 1$, in correspondence with the $\gain$ scheduling case. We observe that the $\gain$ scheduling protocol based on approximate error estimation successfully minimizes the stationary variance to this value. Incorporating $r$ scheduling allows for improving beyond this limit. Panel (c) shows the mean of $\gain$ versus $t$. The horizontal dashed black line corresponds to $\gain=\ell s$ for $s = 1$, corresponding to the value of $\gain$ that minimizes $\sigma^2_\infty$ in the $\gain$ scheduling case. We observe that the $\gain$ scheduling protocol using approximate error estimation successfully converges to $\gain=\ell s $ as desired, without any initial guess for $\gain$. Panel (d) shows the mean of $r$ as a function of $t$, and illustrates that on average, the $r$ scheduling protocol increases $r$ with $t$ to achieve more precise tuning over time. }
\label{Fig:approxerrorschedulingstats}
\end{figure}

Here, we describe approximate error estimation as a practical path forward for scheduling $g$ based on the premise of Eq. (\ref{eq:kschedule}). Equation (\ref{eq:kschedule}) allows for calculating a gain schedule given complete knowledge of the mean and variance of $\Delta\eta_t$ at each step $t$. Here, we assume that we do not have this complete knowledge. Instead, the idea is to substitute $\mu_t$ and $\sigma_t^2$ in Eq. (\ref{eq:kschedule}) with estimated counterparts $\hat{\mu}_t$ and $\hat{\sigma}_t^2$,
\begin{equation}
    \gain_t = \frac{2\hat{\sigma}^2_t s^2}{1-4s^2\hat{\mu}^2_t},
    \label{eq:kscheduleapproximate}
\end{equation}
where $\hat{\mu}_t$ and $\hat{\sigma}_t^2$ can be estimated based on the measurement record $\mathcal{M}$ within a single realization of the IOC protocol, e.g., according to 
\begin{equation}\label{Eq:tildemu}
\hat{\mu}_t = -\frac{\sum_{j=1}^M \left(\sum_{k=1}^N z_{t_{k,j}}\right)}{2MNs}
\end{equation}
and
\begin{equation}\label{Eq:tildesigma}
\hat{\sigma}^2_t = \frac{\sum_{j=1}^M \left(\sum_{k=1}^N z_{t_{k,j}}\right)^2}{4s^2MN^2} - \hat{\mu}_t^2.
\end{equation}
Here, the idea is that at each step $t$, $M$ bins of $N$ samples each are collected into $\mathcal{M} = \{z_i\}_{i=1}^{MN}$ by executing the IOC protocol. The samples are divided into $M$ different bins, indexed by $j$, of $N$ samples each, indexed by $k$, to evaluate Eqs. (\ref{Eq:tildemu}) and (\ref{Eq:tildesigma}), which are then used to update $g$ according to Eq. (\ref{eq:kscheduleapproximate}). This heuristic approach requires knowledge of $s$ and $\mathcal{M}$ only. Updates to $g$ can be made at regular intervals, or even each step $t$ using a sliding window of the prior $MN$ steps.

\begin{figure}
\centering
\includegraphics[width=\columnwidth]{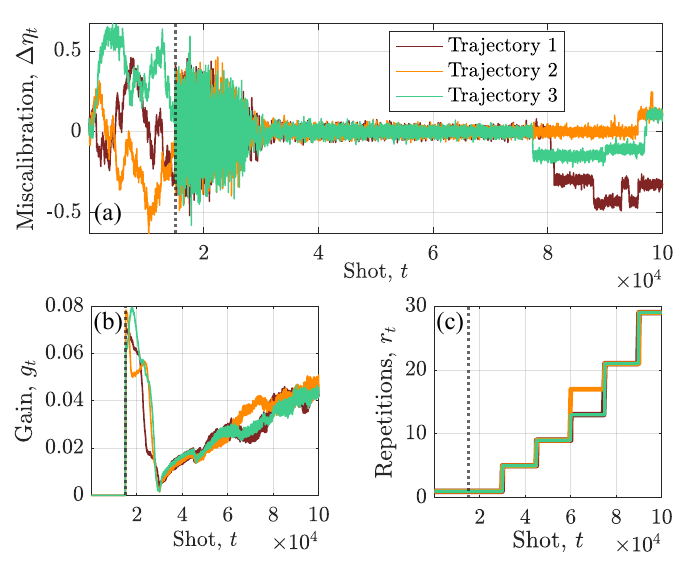} 
\caption{\textbf{Failure modes of the approximate error estimation heuristic.} The performance of three trajectories of a $g,r$-scheduled IOC protocol based on approximate error estimation are shown. All simulation details and hyperparameters are selected to be the same as those associated with the results in Fig. \ref{Fig:approxerrorschedulingstats}, with the exception of $\kappa$, which governs how conservative the $r$ scheduling protocol is. Here, we set $\kappa = 2$ instead of $\kappa = 3$. Panel (a) shows the dynamics of the parameter deviation, $\Delta\eta_t$, as a function of $t$ for each of the three trajectories. Panels (b) and (c) show the associated gain and repetition schedules, respectively. We observe that decreasing $\kappa $ causes a more aggressive repetition schedule for all three trajectories, which leads to a breakdown in performance. This is illustrated in panel (a), where at large values of $t$, $\Delta \eta$ hops into undesired basins and stabilizes itself there, away from its target value of $\Delta\eta=0$. }
\label{Fig:approxerrorschedulingtrajectoriesfailure}
\end{figure}

Repetition scheduling, meanwhile, can be accomplished by first defining a target threshold, $\delta_{max}$, for how much rotation error is tolerable, i.e., where we desire $|\delta_t| \leq \delta_{max}$ for all steps $t$. We can then use the definition $\delta_t = r_t\alpha\Delta\eta_t$ to obtain the following threshold-based repetition schedule, $ r_t \leq \frac{\delta_{max}}{|\alpha\Delta \eta_t |}$, which can be evaluated in practice using the approximate error estimators in Eqs. (\ref{Eq:tildemu}) and (\ref{Eq:tildesigma}) according to
\begin{equation}
     r_t = \frac{\delta_{max}}{|\alpha|(|\hat{\mu}_t| + \kappa \hat{\sigma_t})},
     \label{eq:repetitionscheduling}
\end{equation}
where the denominator stipulates that we are within $\kappa$ estimated standard deviations of the estimated mean. Increasing $\kappa$ or decreasing $\delta_{max}$ leads to more conservative performance. Updates to $r$ via Eq. (\ref{eq:repetitionscheduling}) can be made every $MN$ steps when using the estimators in Eqs. (\ref{Eq:tildemu}) and (\ref{Eq:tildesigma}), because the estimators are functions of $r$ (through their dependence on $s$), which is presumed to be fixed over the interval to do the estimation.

The performance of gain and repetition scheduling based on approximate error estimation is illustrated in Fig. \ref{Fig:approxerrorschedulingstats}, which compares how the mean and variance of the parameter deviation, $\Delta\eta_t$, computed across 50 realizations, behave across three cases: an uncalibrated baseline, the IOC protocol with $g$ scheduling based on approximate error estimation, and the IOC protocol with $g$ and $r$ scheduling based on approximate error estimation. In all cases, the drift of the nominal parameter value, $\eta_{opt,t}$, is described by an unbiased discrete random walk per Eq. (\ref{eq:app:rwupdate}).

In the absence of $r$ scheduling, we can make a clean comparison between the $g$ scheduling case and the analytical derivation in Sec. \ref{App:Sub:IOCBehavior}. We find that $g$ scheduling based on approximate error estimation successfully produces the minimal stationary variance predicted by theory (Fig. \ref{Fig:approxerrorschedulingstats}(b), inset) by automatically converging to the predicted value of $g$ over time ((Fig. \ref{Fig:approxerrorschedulingstats}(c)).

The numerical results in Fig. \ref{Fig:approxerrorschedulingstats} show that these scheduling protocols are able to automatically determine appropriate schedules for $g$ and $r$, even in the absence of any initial guesses. That is, in our numerical experiments we initialize $g_0=0$ and $r_0=1$ to deliberately illustrate that the method can still operate without any tuning of $g_0$ and $r_0$ as additional hyperparameters. In practice, however, seeding the protocol with better initial guesses for $g$ and $r$ can be a more desirable option. In our experience, better choices of $g_0$ and $r_0$ can allow for substantially reducing $M,N$ while still retaining good performance for gain scheduling. However, reducing $M,N$ produces coarser estimates of the mean and variance, and eventually this can cause the scheduling performance to break down.

In practice, the choice of relevant hyperparameters ($M,N,\delta_{max},\kappa$) can substantially impact performance. For example, the use of sampled estimates of $\mu_t$ and $\sigma^2_t$, rather than their exact values, introduces stochastic fluctuations due to sampling noise into the scheduling protocols. It is therefore desirable to strike a balance between the sampling-based fluctuations due to finite $N,M$, which increase in severity as $N,M$ decrease, and the drift-based fluctuations that we aim to track with the IOC protocol in the first place. If the former dominates, then the scheduling protocols are not able to observe the parameter drift beneath the sampling noise floor. This is a failure mode of the scheduling protocol based on approximate error estimation. If the drift dominates, however, then it can be observed and tracked. It is thus desirable to select $N,M$ large enough that the sampling-based fluctuations are small relative to the drift. However, $N,M$ should not be selected to be excessively large, as in addition to carrying a higher memory cost, this allows more drift to occur within the sampling window, leading to higher latency, larger error in the estimates, and deterioration of performance. 

In the absence of $r$ scheduling, as in the light blue curves in Fig. \ref{Fig:approxerrorschedulingstats}, we may treat $r$ as a hyperparameter; if we schedule $r$, we may consider $r_0$ as a hyperparameter. In our numerical experiments, we have observed that if we make $r$ or $r_0$ too small, then for some drift models it is difficult to observe the drift process beneath the sampling-based fluctuations. The resulting failure mode is similar to what we have observed when $N,M$ are selected to be too small. Meanwhile, if $r$ is selected to be too large (or if it becomes too large during the scheduling protocol), this can lead to convergence to an undesired minimum. This particular failure mode is captured in Fig. \ref{Fig:approxerrorschedulingtrajectoriesfailure}, where we observe in panel (a) that $\Delta\eta$ deviates at late times from its target regime of $\Delta\eta\approx 0$, as $r$ becomes too large. This failure is indicative of the IOC protocol converging to the wrong optimum, i.e., due to leaving the linear regime and converging to a root of $\sin(r\alpha\Delta\eta)$ other than $\Delta\eta=0$. Again, it is thus desirable to strike a balance with a choice of $r$ that is large enough to allow for precise tuning, but not so large that it risks driving the protocol to a false minimum.

In practice, we expect that these scheduling heuristics could be valuable in settings where the drift is variable, or where it is desirable to automate the task of gain or repetition selection. However, the techniques are not infallible, and a systematic exploration of how they perform in more complex, real-world drift scenarios would constitute interesting future work. 

\subsubsection{Autocorrelation analysis}
\label{App:SubSub:Autocorrelation}

Another technique for adjusting both $g$ and $r$ that has performed consistently well in simulation is a protocol based on autocorrelation analysis. In this protocol, we initially set the gain $g_0$ to be large and $r_0=1$. We then begin the IOC protocol, keeping a sliding window measurement record of length $h$. At each step when the sliding window is full (i.e., where we have taken $h$ measurement shots since our last update of $g$ or $r$), we compute the autocorrelation over the measurement history window as:
\begin{equation}
    a = \sum_{t=1}^{h-1}z_tz_{t-1}
\end{equation}
There are then four possible cases based on the value of $a$ and chosen hyperparameters $a_{UB}$, $a_{LB}$, $b$ and $r_{max}$:
\begin{enumerate}
    \item If $a>a_{UB}$, increase $g$ and reset the measurement history record.
    \item If $a<a_{LB}<0$, decrease $g$ and reset the measurement history record.
    \item If $|a|<b$ and $r\leq r_{max}$, increase $r$ and reset the measurement history record.
    \item If $b\leq a\leq a_{UB}$ or $a_{LB}\leq a\leq -b$, continue with the current $g,r$ conditions. Note, if one sets $b=a_{UB}=|a_{LB}|$, this condition will never be reached and $g,r$ will be updated every $h$ shots.
\end{enumerate}

This technique for dynamically scheduling $g$ and $r$ is motivated by the fact that a large autocorrelation sum, $a$, indicates that the measurements are likely to be correlated which occurs when the parameter is far from its target, suggesting $g$ should be increased. Meanwhile, a low autocorrelation sum indicates that the measurements are likely to be anticorrelated, which means our update steps are likely causing large oscillations around $\eta_{opt}$, and $g$ should be decreased. Finally, if the autocorrelation sum is near zero, this suggests the measurements are likely to be uncorrelated, as desired.

\begin{figure} 
\centering
\includegraphics[width=\columnwidth]{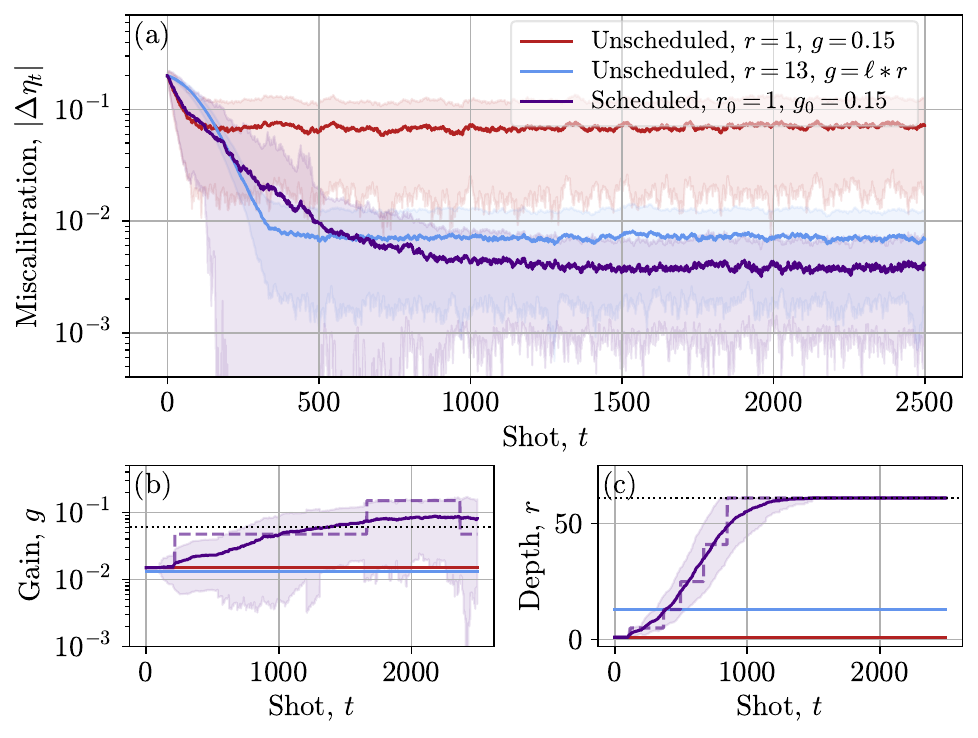}
\caption{\textbf{Scheduling the IOC protocol with autocorrelation analysis.} We compare three simulated executions of the IOC calibration protocol: (1) one execution using constant, naive hyperparameters $g=0.015$ and $r=1$ (red), (2) another with constant, tuned parameters $g=0.013$ and $r=13$ (blue), and (3) a run with dynamically scheduled $g$ and $r$ (indigo) using the autocorrelation analysis heuristic and starting at the same naive initial parameters as the numerical experiment in red. In all cases, we plot the mean and standard deviation of 200 calibration trajectories with a $\Delta\eta_0=0.2$, random walk drift with $\ell=0.001$, per-gate depolarization with $p=0.001$, and a depolarizing SPAM error channel with $p_{SPAM}=0.01$. Over time, the heuristic raises $r$ from its initial value of $r_0=1$ to its maximum allowed value of $61$ as the average $\Delta\eta$ lowers, and $g$ fluctuates and gradually raises in response to the measured data. The average $g$ approaches $g=\ell r_{max}=0.061$, demonstrating that the heuristic ``learns'' the approximate value of $\ell$. The $g$ and $r$ for a single trajectory are shown in a dashed line to demonstrate the step changes the heuristic makes over time, with the variance of the $r$ plot capped due to fixing $r_{max}=61$. By scheduling the IOC protocol's gain and circuit repetitions in this way, we achieve fast initial convergence and lower stationary variance than the unscheduled cases without assuming any prior knowledge of the drift characteristics. Note, the feature in the indigo shading in (a) between 200 and 1000 shots is deceivingly low due to a combination of a small mean value and a large variance in that range. Most scheduled calibration trajectories track closely with the plotted mean.}
\label{Fig:AppAutocorrSched}
\end{figure}

Figure \ref{Fig:AppAutocorrSched} shows the performance of this scheduling protocol compared to two unscheduled baselines with different $r$ and $g$ settings: first with $r=1$ and a naively set $g=0.015$, and second with $r=13$ and $g=\ell s=0.013$ (where the latter matches the settings used in Figs. \ref{Fig:IOC} and \ref{Fig:MultiParameterGxGy}, where $r=13$ was determined in simulation to be the highest value of $r$ which could correct for an initial offset $\Delta\eta_0=0.2$ without converging to the wrong optimization minimum). By introducing the autocorrelation scheduling technique, we observe significantly accelerated transient convergence as well as an improvement in stationary variance over the unscheduled cases, while requiring no prior knowledge of the behavior of the drift process. This protocol can operate with less computation overhead than our prior heuristic, which could lead to more lightweight implementations suitable for FPGAs \cite{IOCImplementation}.

As in the scheduling protocol based on approximate error estimation, the performance can depend strongly on the choices of hyperparameter values. For these numerical simulations, we choose $h = 100$, $a_{UB} = 20$, $a_{LB} = -20$, $b = 1$ and $r_{max}=61$. The update rules for increasing or decreasing $g$ are given by $g_i = \sqrt{10}g_{i-1}$ and $g_i = \frac{g_{i-1}}{\sqrt{10}}$, respectively, and the update rule for increasing $r$ was given by $r_i = r_{i-1}+4n$ with $n\in\{1,2,..\}$ such that $r$ follows the sequence $\{1,5,13,25,41,61\}$. The rates of change for $g$ and $r$ were chosen heuristically from our simulations, with $r$ increasing quickly for a visibly improved convergence rate and $g$ changing on a logarithmic scale so that the calibration would converge to the right order of magnitude of the drift parameter $\ell$ from a naive initial guess. With $r_{max}=61$ in this simulation, we know by Eq. \eqref{eq:optkinfty} that the gain which will minimize stationary variance is $g=0.061$, which $g$ approaches in Fig. \ref{Fig:AppAutocorrSched}(b). Thus, we observe that this heuristic learns the approximate value of $\ell$ by the convergence behavior of its gain.

\subsection{DOC protocol scheduling heuristic}
\label{App:Sub:DOCHeuristic}

\begin{figure}
\centering
\includegraphics[width=\columnwidth]{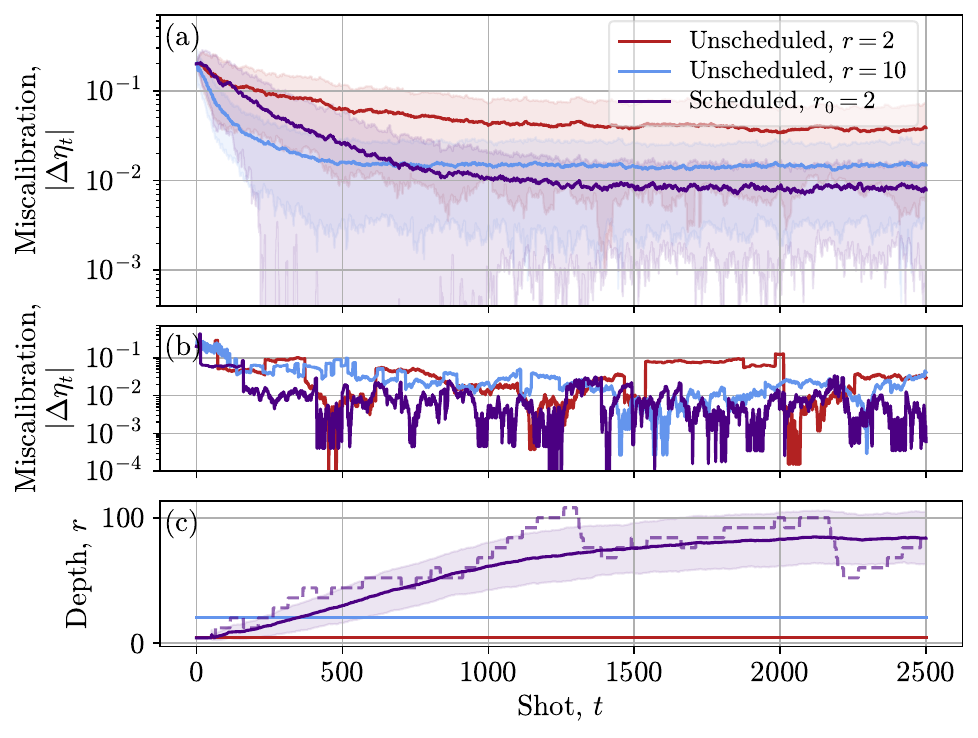}
\caption{\textbf{Demonstrating heuristic scheduling of the DOC calibration protocol.} We compare three executions of the DOC calibration protocol with different hyperparameter choices: (1) constant, minimum $r=2$ (red), (2) constant, experimentally informed $r=10$ (blue), and (3) scheduled $r$ using the heuristic scheduling algorithm in Sec.~\ref{App:Sub:DOCHeuristic} and starting at the minimum $r_0=2$ (indigo). In all cases, we plot the mean and standard deviation of 200 calibration trajectories with $\Delta\eta_0=0.2$, random walk drift with $\ell=0.001$, per-gate depolarization with $p=0.001$, and a depolarizing SPAM error with $p_{SPAM}=0.01$. We see that over time, the heuristic scheduling protocol raises and lowers $r$ as the average wait time between observed errors changes. The value of $r$ hits a soft upper bound enforced by the point at which errors from SPAM and depolarization outweigh the error from stochastic parameter drift. A single calibration trajectory for each experimental condition is shown in (b) and a dashed line representing $r$ for the scheduled calibration of that trajectory is shown in (c) to demonstrate the step changes the scheduling heuristic makes over time. By scheduling the DOC protocol's circuit repetitions in this way, we achieve fast initial convergence and lower stationary variance than the unscheduled cases without assuming any prior knowledge of the noise characteristics. Without adaptivity, the number of shots required between parameter updates increases as the calibration error reduces, making subsequent calibration updates significantly more expensive. Note, the feature in the indigo shading in (a) between approximately 250 and 1000 shots is deceivingly low due to a combination of a small mean value and a large variance in that range.}
\label{Fig:DOCadaptive}
\end{figure}

Here, we present a heuristic approach for tuning $r$ in the DOC protocol that has demonstrated strong performance in simulation. This approach operates as follows:
\begin{enumerate}
    \item Choose the cutoff parameter, $n$. Here, we choose $n=2$ as a sufficiently small number to be able to update quickly, but \mbox{$n>1$} so that Eq. \eqref{Eq:MVUE} never evaluates to $0$.
    \item Introduce an upper bound $N_{\rm max}$ on the number of shots to take per episode. In our examples, we choose $N_{\rm max} = 50$ shots.
    \item For each episode, if we require fewer than $N_{\rm max}$ shots to observe $n$ failures, then update $\eta$ using the DOC rule in Eq.~\eqref{eq:DOCupdate}.
    \begin{enumerate}
        \item If we observed $n$ failures in fewer than $N_{min}$ shots, the circuit is failing too frequently, and so we reduce the circuit depth using the heuristic $r_{t+1} = r_t-8$. Here, we choose $N_{min}=10$ shots.
    \end{enumerate}
    \item If we do not observe $n$ failures before reaching $N_{\rm{max}}$ shots, then stop the episode, do not adjust the parameter, and start a new episode increasing the circuit depth as $r_{t+1}=r_t+8$.
\end{enumerate}
Fig.~\ref{Fig:DOCadaptive} compares the adaptive and non-adaptive DOC approaches when calibrating a $\gx$ gate in the presence of a random walk parameter drift, SPAM error, and per-gate depolarization error.

We note that tuning the upper and lower bounds of the DOC circuit repetition scheduling method can aid in improving convergence rates and results. In this work, we primarily use the bounds $N_{min}=10$, $N_{max}=50$, and $n=2$. We have found that having a smaller $N_{max}$ can cause the results to converge faster. It is also important to have $N_{min}>0$ in order to reduce $r$ if it becomes too large, as $N_{min}=0$ can result in diverging behavior beyond initial tuning. We also note that the rate of change of $r$ is heuristically chosen here, and we have found that other heuristics, such as halving or doubling $r$ based on the same $N_{min}$ and $N_{max}$, can be similarly effective.

\section{Robustness to different drift processes}
\label{App:Noise}

We have explored the performance of IOC and DOC protocols in different drifting settings. In this section, we display results for cases where drift in $\etaopt$ is described by (1) a discrete random walk, per Eq. (\ref{eq:app:rwupdate}), (2) an Ornstein-Uhlenbeck process compounded by a jump process, and (3) a $1/f$ noise process. 

\begin{figure*} 
\centering
\includegraphics[width=\textwidth]{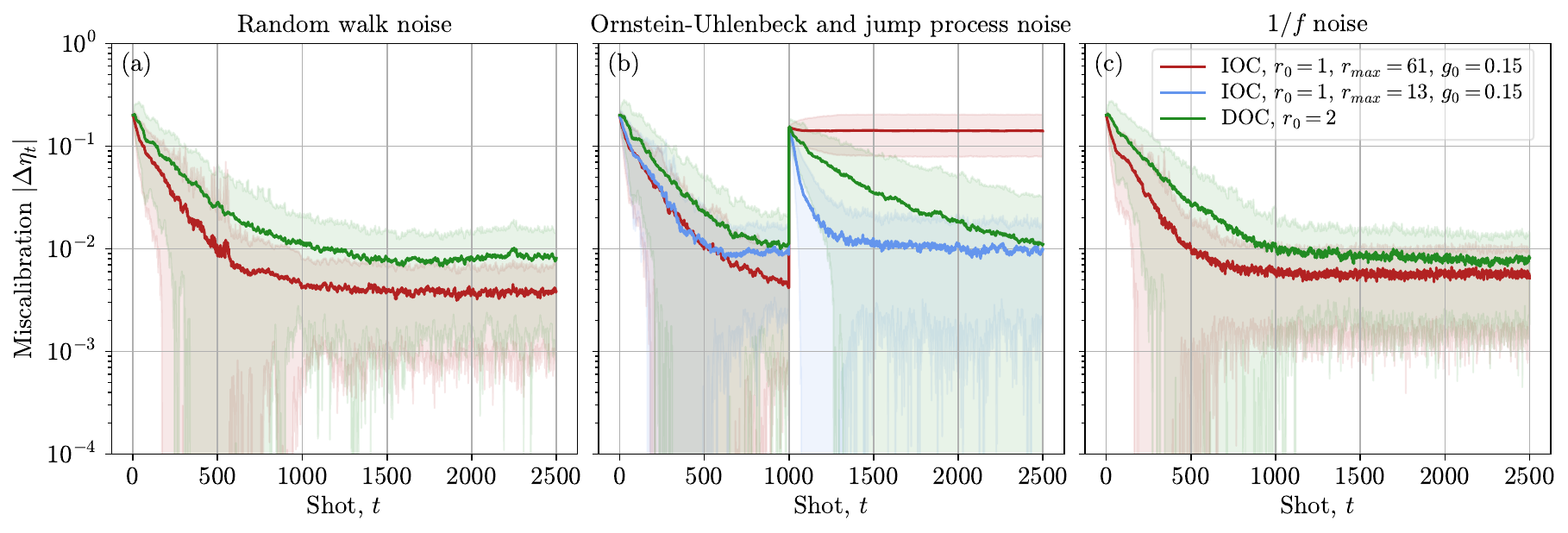}
\caption{\textbf{Demonstrating the performance of IOC and DOC calibration protocols with different drift processes.} We demonstrate the performance of both the IOC and DOC calibration protocols in responding to three types of drift processes: (a) a random walk, (b) an Ornstein-Uhlenbeck process compounded by a jump process at shot number 1000, and (c) a $1/f$ noise process. The parameters describing each stochastic noise process are further described in Sec. \ref{App:Noise}. The figure plots the mean results over 200 trajectories following the scheduling heuristics described in Sections \ref{App:SubSub:Autocorrelation} (red) and \ref{App:Sub:DOCHeuristic} (green). In (b) we add a modified version of the IOC autocorrelation analysis heuristic in which $r_{max}=13$ (blue), to demonstrate the importance of $r_{max}$ in the IOC protocol for robustness to stochastic jump processes. In all simulations, we also include a per-gate depolarization with $p=0.001$, a depolarizing SPAM error channel with $p_{SPAM}=0.01$, and $\Delta\eta_0=0.2$.}
\label{Fig:AppNoiseModels}
\end{figure*}

The Ornstein-Uhlenbeck process is modeled as:
\begin{equation}
    \eta_{opt,t} = \eta_{opt,t-1}e^{-\alpha}+\sigma\epsilon_t
\end{equation}\label{eq:ouprocess}

\noindent where we start $\eta_{opt,0}$ at some initial error offset, $\alpha$ is the mean reversion coefficient, $\sigma$ is the volatility coefficient and $\epsilon_t$ is drawn from $\mathcal{N}(0,1)$. In the trials shown in this section, we set $\alpha=0.0001$ and $\sigma=0.001$. We also add a stochastic jump process to study the resilience of our calibration protocols to sudden parameter changes, where $\etaopt$ changes by $0.15$ at shot number 1000.

The $1/f$ noise process is simulated as
\begin{equation}
    \eta_{opt,t} = \sum_{i=1}^{7}k(\eta_{opt,t-1}e^{-\alpha}+\sigma\epsilon_t),
\end{equation}\label{eq:1fprocess}

\noindent where $k=0.001$ is a scaling factor for a sum of multiple Ornstein-Uhlenbeck processes with mean reversion coefficients $\alpha=10*\frac{1}{4}^i$ dependent on their index $i$, volatility coefficients $\sigma=2^i(1-e^{-2*\alpha})$, and $\epsilon_t$ drawn from $\mathcal{N}(0,1)$.

Figure \ref{Fig:AppNoiseModels} shows the average performance over 200 trajectories of the IOC and DOC calibration protocol with each type of drift process, a SPAM depolarization error channel with $p_{SPAM}=0.01$, and a per-gate depolarization channel with $p=0.001$. The figure compares the calibration performance for both protocols following the heuristics described in Sections \ref{App:SubSub:Autocorrelation} and \ref{App:Sub:DOCHeuristic}.

For each of the three drift processes, the IOC protocol tends to converge faster than the DOC protocol. However, the IOC protocol can be less robust to jump processes, as shown in Fig. \ref{Fig:AppNoiseModels}(b). This is because the scheduling heuristic for the DOC protocol has a built-in mechanism to decrease $r$ when the mean is detected to be far from its optimal. However, the autocorrelation analysis heuristic of the IOC protocol does not, and thus the $r_{max}=61$ setting is too high to remain robust to a sudden jump in $\etaopt$. This jump can push the calibration into the wrong minimum of the optimization landscape, and thus $\eta$ may not properly converge after the jump occurs (in this example, see behavior of red curve in panel (b) for a jump of size $j=0.15$ given $r_{max}=61$). 

To minimize this effect, we observe that the IOC protocol will converge on average for $|\Delta\eta|<\pi/r$, and thus the protocol should be robust to jumps, on average, if $|\Delta\eta_t+j|<\pi/r$ for a jump occurring at time $t$. We can assume $\Delta\eta_t$ has been made small by calibration, and thus set $r_{max}<\pi/|j|$ based on the maximum jump magnitude to which we wish to be robust. In the trial in Fig. \ref{Fig:AppNoiseModels}(b), $r_{max}=13$ provides robustness to jumps in $\etaopt$ of magnitude $|j|<0.242$ given the jump occurs when $\Delta\eta$ is close to $0$. This is shown by the blue $r_{max}=13$ curve recovering to a well-calibrated state after the jump occurs.

We have also experimented with alternate methods to catch and correct jump processes in the IOC protocol, but this $r_{max}$ setting is one of the most efficient. Other methods we have explored include periodic or small-sample mean estimate-informed resets or reductions of $r$, but these can tend to add artificial noise when jumps are not present.

\end{document}